\documentclass[10pt,journal,compsoc]{IEEEtran}
  \usepackage{cite}
\ifCLASSINFOpdf
   \usepackage[pdftex]{graphicx}
   \graphicspath{{../pdf/}{../jpeg/}}
   \DeclareGraphicsExtensions{.pdf,.jpeg,.png}
\else
\fi
\usepackage{filecontents}
\usepackage{tikz}

\usepackage{lipsum}
\usepackage{fancyhdr}
\usepackage{epsfig,endnotes,color}
\usepackage{graphicx}
\usepackage{amsmath,bm}
\usepackage{graphicx}
\usepackage{url}
\usepackage{placeins}
\usepackage{CJKutf8}
\usepackage{subfigure}
\usepackage{rotating} 
\usepackage{lscape, latexsym, amssymb, multirow}
\usepackage{multirow}
\usepackage{mathtools, bbm, color}
\usepackage{booktabs}
\usepackage{rotating}
\usepackage{amsthm,mathrsfs,amsfonts,dsfont}

\usepackage{enumerate}
\usepackage{hhline}
\usepackage{enumitem}
\usepackage{tabu}
\usepackage{cite}
\usepackage{colortbl} 
\usepackage{enumerate}
\usepackage{caption}
\usepackage{footnote}
\usepackage{color}
\usepackage{colortbl}
\usepackage{pifont}
\usepackage{bbding}
\usepackage{threeparttable}
\usepackage[framemethod=TikZ]{mdframed}
\usepackage{etoolbox,xspace}
\usepackage{algorithm}
\usepackage{algpseudocode}
\usepackage{bookmark}
\usepackage{longtable}
\usepackage{supertabular}
\usepackage{titlesec}
\usepackage{siunitx}
\usepackage{amssymb}
\usepackage{hyperref}
\usepackage{listings}
\usepackage{pifont}
\usepackage[perpage,symbol*]{footmisc}
\DefineFNsymbols{circled}{{\ding{192}}{\ding{193}}{\ding{194}}
	{\ding{195}}{\ding{196}}{\ding{197}}{\ding{198}}{\ding{199}}{\ding{200}}{\ding{201}}}
\setfnsymbol{circled}

\usepackage{color} 

\newcommand{\wrpdeleteY}[1]{}

\newcommand{\sys}{{G-Fuzz}\xspace}
\newcommand{\gv}{{gVisor}\xspace}
\newcommand{\aflgo}{{AFLGo}\xspace}

\begin{document}
%
\title{\sys: A Directed Fuzzing Framework for \gv}
%
%
%
%

\author{Yuwei~Li,  
        Yuan~Chen,   
        Shouling~Ji,    
        Xuhong~Zhang,   
        Guanglu~Yan,    
        Alex X.~Liu,    \\
        Chunming~Wu,   
        Zulie~Pan,   
        Peng~Lin
\IEEEcompsocitemizethanks{\IEEEcompsocthanksitem 
S.~Ji and C.~Wu are with the corresponding authors.
Y.~Li and Y.~Chen are the co-first authors.

\IEEEcompsocthanksitem 

Y.~Li is with the College of Electronic Engineering, National University of Defense Technology, Hefei, Anhui, 230000, China. 
Part of this work was down when she was a Ph.D student at the College of Computer Science and Technology, Zhejiang University, Hangzhou, Zhejiang, 310027, China.
Email: liyuwei@nudt.edu.cn


\IEEEcompsocthanksitem 
S.~Ji is with the College of Computer Science and Technology, Zhejiang University, Hangzhou, Zhejiang, 310027, China.
Email: sji@zju.edu.cn

\IEEEcompsocthanksitem 
X.~Zhang is with the School of Software Technology, Zhejiang University, Hangzhou, Zhejiang, 310027, China.
Email: zhangxuhong@zju.edu.cn

\IEEEcompsocthanksitem 
G.~Yan and A.~Liu are with Ant Group, Hangzhou, Zhejiang, 310063, China. 
Email: flankreader@gmail.com, alexliu@antgroup.com

\IEEEcompsocthanksitem 
Y.~Chen and C.~Wu are with College of Computer Science and Technology, Zhejiang University, Hangzhou, Zhejiang, 310027, China.
Email: chenyuan@zju.edu.cn, wuchunming@zju.edu.cn

\IEEEcompsocthanksitem 
Z.~Pan is with the College of Electronic Engineering, National University of Defense Technology, Hefei, Anhui, 230000, China.
Email: panzulie17@nudt.edu.cn

\IEEEcompsocthanksitem 
P.~Lin is with Chinese Aeronautical Establishment and 
AVIC Artificial Intelligence Research Institute, Beijing, 100029, China.
Email: 13940001294@163.com

}
}

%
%

\markboth{IEEE TRANSACTIONS ON DEPENDABLE AND SECURE COMPUTING,~Vol.~21, No.~1,  2024}%
{Shell \MakeLowercase{\textit{et al.}}: Bare Demo of IEEEtran.cls for Computer Society Journals}
%



\IEEEtitleabstractindextext{%
\begin{abstract}
gVisor is a Google-published application-level kernel for containers. As gVisor is lightweight and has sound isolation, it has been widely used in many IT enterprises~\cite{Stripe, DigitalOcean, Cloundflare}.
When a new vulnerability of the upstream gVisor is found, it is important for the downstream developers to test the corresponding code to maintain the security. To achieve this aim, directed fuzzing is promising. Nevertheless, there are many challenges in applying existing directed fuzzing methods for gVisor. The core reason is that existing directed fuzzers are mainly for general C/C++ applications, while gVisor is an OS kernel written in the Go language. To address the above challenges, we propose G-Fuzz, a directed fuzzing framework for gVisor. There are three core methods in G-Fuzz, including lightweight and fine-grained distance calculation, target related syscall inference and utilization, and exploration and exploitation dynamic switch. Note that the methods of G-Fuzz are general and can be transferred to other OS kernels. We conduct extensive experiments to evaluate the performance of G-Fuzz. Compared to Syzkaller, the state-of-the-art kernel fuzzer, G-Fuzz outperforms it significantly. Furthermore, we have rigorously evaluated the importance for each core method of G-Fuzz. G-Fuzz has been deployed in industry and has detected multiple serious vulnerabilities.
\end{abstract}

}

\maketitle

\IEEEdisplaynontitleabstractindextext

%
\IEEEpeerreviewmaketitle

\section{Introduction}\label{sec:intro}
\gv~\cite{gVisor} is an application kernel that aims to provide secure isolation between the host kernel and the applications running inside the containers.
Compared to the virtual machine, \gv is more lightweight and can provide a similar isolation level.
Thus, it has been adopted in many IT companies~\cite{Stripe, DigitalOcean, Cloundflare}.

The vulnerabilities of \gv may severely impact the security and stability of the production environments.
Those IT companies usually implement a customized \gv based on the specific application scenario, and they may face many security issues.
How to test whether the modified code of the customized \gv introduces new vulnerabilities?
When a vulnerability of \gv has been patched, whether the patch can fix the vulnerability?
To address the above questions, directed fuzzing is one promising and practical solution.
Directed fuzzing aims at generating the test cases that can trigger the specified target code.
The current directed fuzzing techniques can be mainly categorized into two classes: symbolic execution based whitebox directed fuzzing~\cite{marinescu2013katch, jin2012bugredux} and directed greybox fuzzing (DGF)~\cite{bohme2017directed, chen2018hawkeye}.
As symbolic execution has limitations such as path explosion and complex constraints, it is hard to apply symbolic execution based directed fuzzing on large and complex software like kernels.
For OS kernels, compared to symbolic execution based methods, DGF has better scalability and is more feasible.
Nevertheless, there are still many challenges when applying existing DGF techniques on \gv as follows.

\textbf{High Time Overhead in Distance Calculation.}
DGF usually leverages distance information as the guidance for testing the target.
Specifically, DGF selects the closer inputs as seeds to generate new inputs.
For each code (e.g., function/basic block) of the test program, DGF calculates its static distance to the target function/basic block before fuzzing.
Then, during the fuzzing process, the executed inputs that cover the code closer to the target will be selected as the seeds to further generate new inputs.
However, the time overhead of the current DGF in calculating static distance is hardly affordable for large software or systems like OS kernels.
For instance,  when leveraging the state-of-the-art DGF method, \aflgo~\cite{bohme2017directed},  to calculate the static distance for \gv, the average time it takes is more than 16 hours.
Therefore, the time burden brought by the current static distance methods severely impedes the efficiency of directed fuzzing.

\textbf{The Limitations of the Distance Information.}
Even with the distance information, there are still many challenges.
First, the distance information may not be precise.
For instance, when constructing a program's function Call Graph (CG), \aflgo does not consider the indirect calls.
As a result, the CG may lose some critical edges, which further affects the precision of the distance information.
In addition, both \aflgo and Hawkeye~\cite{chen2018hawkeye}, the state-of-the-art directed greybox fuzzers, approximate basic block level distance by multiplying function level distance with a constant (e.g., 10 in AFLGo paper~\cite{bohme2017directed}), which may cause many biases.
Second, although the distance information can provide guidance in triggering the target, it still has disadvantages.
The paths that are closer to the target may not necessarily be easier to get there, as distance is spatial information and may not provide an equivalent measure of the difficulty in reaching the target.
Moreover, in patch testing, a typical application scenario of directed fuzzing, there may be multiple paths leading to the patch target. 
Therefore, only focusing on the shortest path might not be able to comprehensively test the patch code.
Thus, to improve DGF, it is necessary to improve the precision of the distance calculation and make reasonable use of the distance information.

\textbf{Difference of Inputs.}
Existing DGF mainly focuses on testing the general user-space applications that take files as inputs.
Although \gv runs in the user mode, in essence, it is an OS kernel that takes  syscall\footnote{We use syscall to indicate system call in this paper.} sequences as inputs.
Compared to files, the structure and semantic requirements for syscall sequences are more strict.
When testing kernels, it is necessary to provide meaningful syscall sequences as inputs. 
Otherwise, only the shallow code of the kernel can be triggered.
\gv implements more than 200 Linux syscalls, and each syscall may have several parameters.
As a result, the whole input space for fuzzing \gv is extremely vast.
However, for directed fuzzing, the target is triggered by a limited syscalls.
Thus, identifying the target related syscalls to reduce the input space needed to be explored is crucial to directed fuzzing for kernels.

To address the above challenges, we propose \sys, a directed fuzzing framework for \gv, including three principal methods.

\textbf{Lightweight and Fine-grained Distance Calculation.}
To solve the problem of the high time overhead and low precision in distance calculation, we propose a \emph{lightweight and fine-grained distance calculation} method.
First, we perform \emph{reachability analysis} to find the paths that can lead to the target.
We then calculate the distance only for the code on these reachable paths instead of all code.
Second, we utilize the \emph{Breadth First Search (BFS)}, a less complex algorithm than the Dijkstra algorithm~\cite{dijkstra} that used by existing DGF~\cite{bohme2017directed, chen2018hawkeye}, to calculate the distance between two nodes.
To solve the false negatives, we use type analysis to identify the indirect calls.
Then, we construct an inter-procedural CFG to calculate basic block level distance, which is more fine-grained than existing DGF's distance.

\textbf{Target Related Syscall Inference and Utilization.}
To reduce the input space that needs to be explored, we propose a \emph{target related syscall inference and utilization} method.
Before directed fuzzing, \sys automatically infers which syscalls are related to the specified target.
Specifically, we propose eight inference rules based on static analysis and the expert knowledge of the \gv code.
Then, based on the inference results, \sys utilizes the inferred syscalls in the mutation operations to generate new inputs during the fuzzing process.
To make efficient use of the inferred syscalls, we adjust the selection probability of each syscall according to the dynamic information of the fuzzing process.
In addition, considering the dependencies of different syscalls, we adjust the order of each syscall in the mutation process to improve the semantic correctness of the generated inputs.

\textbf{Exploration and Exploitation Dynamic Switch.}
To make reasonable use of the distance information, we propose the \emph{exploration and exploitation dynamic switch} method.
There are two modes of \sys in the fuzzing process: \emph{exploration} and \emph{exploitation}.
In the \emph{exploration} mode, \sys acts like a ``coverage-based" fuzzer,  aiming at covering as many paths as possible, which can increase the diversity of the seeds and mitigate false negatives (e.g., missing CG edges).
In the \emph{exploitation} mode, \sys pays more attention to the target by selecting the closer seeds, which aims at accelerating triggering the target.
Exploring more paths may increase the probability of triggering the target, but it is not efficient.
Only focusing on testing the closer paths may make the fuzzing fall into local optimal.
Therefore, it is necessary to make a trade-off and an adaptive adjustment.
\sys adopts a dynamic strategy to adjust the mode selection according to the feedback from the current fuzzing state.
During the fuzzing process, if one mode does not make any progress over a time threshold, \sys will switch to the other mode adaptively.

To evaluate \sys, we test its performance in three typical application scenarios of directed fuzzing, including general target testing, patch testing and bug reproduction.
We compare \sys to Syzkaller and Syz-Go.
Syzkaller is a state-of-the-art kernel fuzzer, and Syz-Go is implemented by us by applying the advanced directed fuzzing methods of AFLGo on Syzkaller.
The experimental results demonstrate that \sys achieves more efficient and stable performance than both Syzkaller and Syz-Go.  
Out of the evaluated 59 typical targets, \sys outperforms Syzkaller on 58 ones, and outperforms Syz-Go on 53 ones.
\sys achieves at least twice speed in trigger the targets than Syzkaller and Syz-Go on 32 targets.
On 5 targets, only \sys can successfully trigger while Syzkaller cannot trigger.
On 2 targets, only \sys can successfully trigger while Syz-Go cannot trigger.
We also evaluate the effectiveness of each core method of \sys separately in \S~\ref{sec:further}, and the results show that all methods make important contribution to the outstanding performance of \sys.
Furthermore, we have deployed \sys in Ant Group, a world-leading IT company, and \sys has detected multiple real-world vulnerabilities of its customized \gv.

The contributions of our paper are summarized as follows. 

\begin{itemize}
\item \textbf{Novel and General Methods.} 
We propose \sys, a directed fuzzing framework for \gv, with three novel methods: \emph{lightweight and fine-grained distance calculation}, \emph{target related syscall inference and utilization} and \emph{exploration and exploitation dynamic switch}.
The methods of \sys are general and scalable, which can be easily extended to testing more OS kernels.
Based on the proposed methods of G-Fuzz, we also implement a prototype called G-Fuzz-Linux, for directed fuzzing Linux kernels.

\item \textbf{Significant Performance.} 
We conduct extensive experiments to evaluate the performance of \sys.
The experimental results demonstrate that \sys outperforms the state-of-the-art kernel fuzzer significantly.

\item \textbf{Real-world Impacts.} We have applied \sys in practice and leverage it to discover multiple real-world vulnerabilities of the downstream \gv in industry.
To facilitate further research on OS kernel directed fuzzing, we have open-sourced  both \sys~\cite{gfuzz-platform} and G-Fuzz-Linux~\cite{gfuzz-linux}.

\end{itemize}

\section{Background}

In this section, we first give a brief introduction about \gv. 
Then, we introduce directed fuzzing techniques.

\subsection{gVisor}

\begin{figure}[tbp]
    \centering
    \includegraphics[width=3.6in]{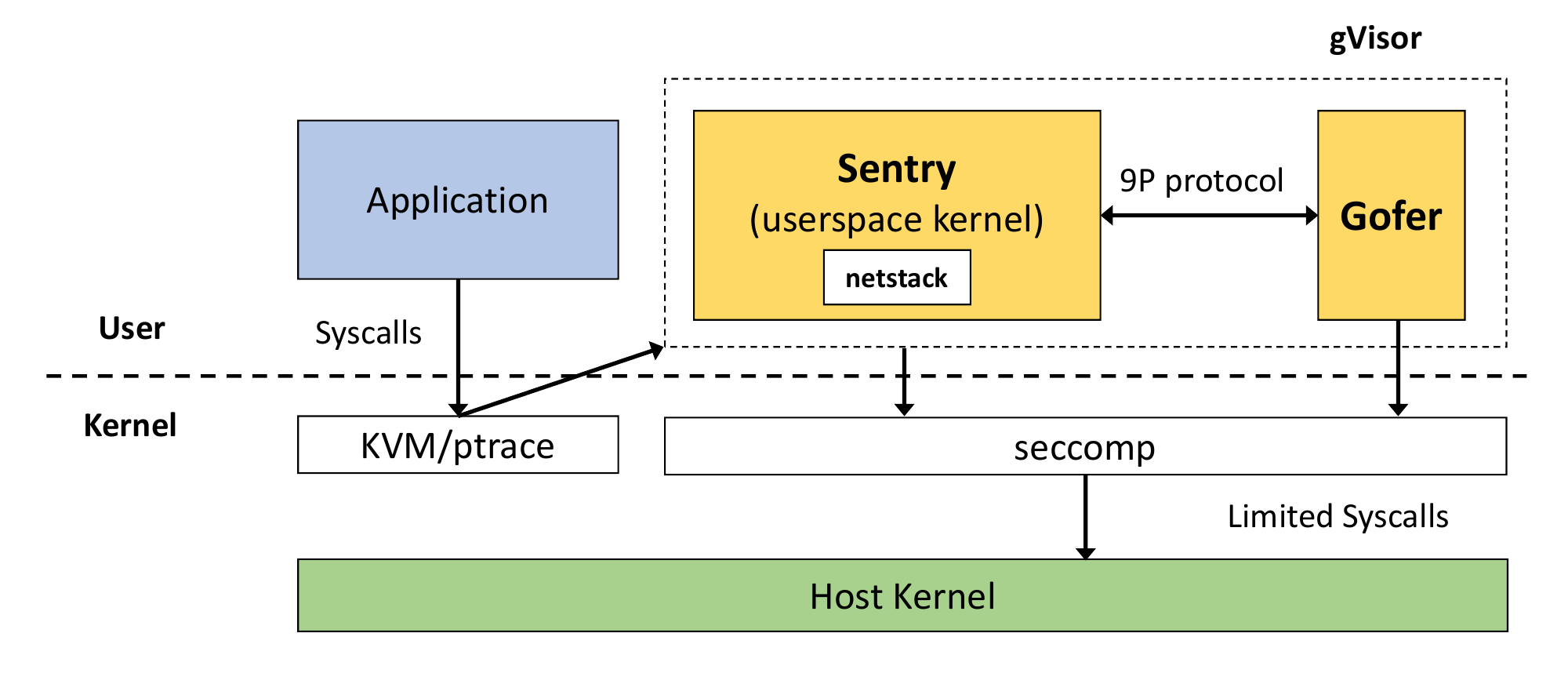}
    \caption{The framework of \gv.}
    \vspace{-1mm}
    \label{fig:gVisor}
\end{figure}

\gv is an application kernel, which is mainly used in virtual environments such as containers.
It implements the isolation by intercepting the syscalls requested from the applications in containers, acting as a guest kernel.
\gv supports most Linux syscalls and runs as a standard, unprivileged process in user-space.
Specifically,  \gv supports 260 syscalls for the AMD64 arch in version \texttt{release-20210125.0}. 
The yellow color part of Fig.~\ref{fig:gVisor} presents the framework of \gv. 
It consists of two primary components: Sentry and Gofer, which run as two processes.
Sentry is the core of \gv, which is responsible for processing the syscalls from the user-space applications.
\gv provides two modes: KVM and ptrace, to redirect the syscalls to Sentry.
To provide extra security protection, \gv adopts the \texttt{seccomp (secure computing mode)} mechanism to restrict the available syscalls to the host kernel.
For instance, the Sentry process has no access to file-related syscalls.
Therefore, when Sentry needs to read or write files on the host file system, it will communicate with the Gofer process using the 9P protocol.
Gofer mediates all these host file-system accesses, providing an additional level of isolation.
\gv also implements a user-space network stack (i.e., \texttt{netstack}), which can process most network-related tasks.

\vspace*{-2mm}

\subsection{Directed Fuzzing}\label{subsec:directfuzz}
Given a target site of a program, the goal of directed fuzzing is to generate the inputs that can trigger it. 
Compared to coverage-based fuzzing, directed fuzzing is faster in detecting vulnerabilities that locate in specified sites.
The original directed fuzzing methods are symbolic execution based~\cite{person2011directed,marinescu2013katch,Ganesh2009Taint,ma2011directed}, which cast the reachability problem as the iterative constraint satisfaction problem~\cite{bohme2017directed}.
Nevertheless, due to the heavyweight program analysis and the difficulty of constraint solving, these methods suffer from the issues of lousy scalability.

To solve these issues, DGF techniques~\cite{bohme2017directed,chen2018hawkeye} are proposed, which leverage the distance information to guide the fuzzing in generating inputs that can reach the target.
Before the fuzzing, DGF calculates a static distance for each component (e.g., basic block or function) of a program to the target.
During the fuzzing process, the inputs that are closer to the target are selected as seeds and given more mutation times.
The distance of an input is calculated based on the static distance of its covered paths.
Compared to symbolic execution based directed fuzzing methods, DGF is more practical and has better performance.
For instance,  \aflgo, a state-of-the-art DGF, spends less than 20 minutes triggering the \emph{heartbleed} vulnerability, while KATCH~\cite{marinescu2013katch}, a state-of-the-art symbolic execution based fuzzer, cannot trigger this vulnerability within 24 hours~\cite{bohme2017directed}.
Therefore, in this paper, we choose to leverage the DGF method for testing \gv. 
There are many works to improve the efficiency of DGF~\cite{ye2020rdfuzz,wang2016seededfuzz,zong2020fuzzguard, lee2021constraint,you2017semfuzz}.
Nevertheless, the above directed fuzzing work is still not appropriate for testing OS kernels.
As we discuss in \S\ref{sec:intro}, there are many challenges in applying existing DGF techniques on kernels.
That is what we aim to solve in this paper.

\section{Design of \sys}

In this section, we first present the overview of \sys. Then we describe each core method of \sys in detail.

\begin{figure*}[tbp]
    \centering
    \includegraphics[width=6in]{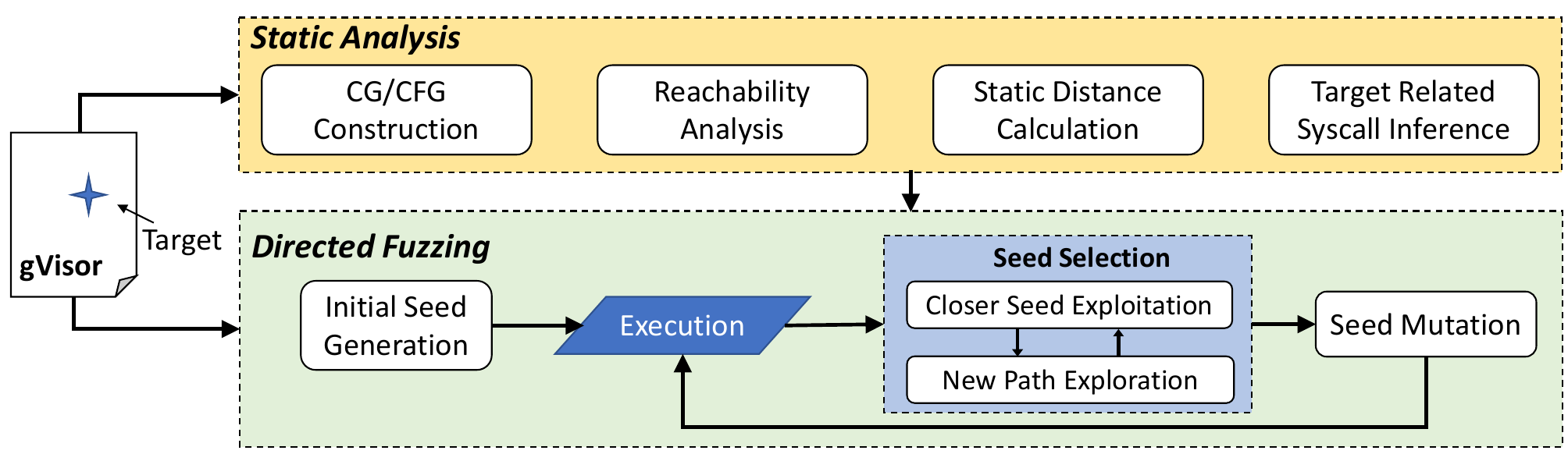}
    \caption{The overview of \sys.}
     \vspace*{-5mm}
    \label{fig:gfuzz-overview}
\end{figure*}

\subsection{Overview}

Fig.~\ref{fig:gfuzz-overview} illustrates the overview of \sys, which consists of two main components: static analysis and directed fuzzing.
Given a target in \gv, the static analysis component aims to extract the information for directed fuzzing, including static distance, target related syscalls, etc.
Then, the directed fuzzing component leverages the extracted information to steer the fuzzer in generating inputs that can trigger the target.

\textbf{Static Analysis.}
There are four main steps involved in the static analysis component of \sys.
First, \sys constructs the CG and CFG of \gv. 
Second, based on the CG and CFG, \sys performs \emph{reachability analysis} for the given target to find the paths in \gv which can lead to the target.
We name the code on these paths as the \emph{reachable set} of the target.
Third, \sys calculates the static distance for each basic block in the \emph{reachable set}.
Finally, \sys determines which syscalls are related to the target based on the automatic inference rules, which are designed based on static analysis and expert knowledge. 
In addition, \sys leverages the static distance information to measure the relevance between each inferred syscall with the target.

\textbf{Directed Fuzzing.}
After static analysis, \sys moves on to directed fuzzing.
The initial seeds are constructed based on the inferred target related syscalls.
Next, \sys selects the inputs with good performance as seeds throughout the fuzzing process.
There are two strategies for selecting inputs. One is selecting inputs that find new paths (i.e., \emph{exploration}  strategy), and the other is selecting inputs closer to the target (i.e., \emph{exploitation} strategy).
To make reasonable use of the distance information and alleviate its limitations, we propose an \emph{exploration and exploitation dynamic switch} method to seek a balance.
When the fuzzer gets ``stuck" by using a seed selection strategy over a time threshold, \sys will switch to the other strategy adaptively.

As a mutation-based fuzzer, \sys mutates the seeds to generate new inputs.
Considering that \gv is an OS kernel, whose inputs are syscall sequences, we propose the following methods in \sys to generate high-quality inputs for directed fuzzing.
First, \sys leverages the inferred syscalls in the mutation process to generate inputs that tend  to trigger the target.
Second, as there may exist false positives in the inferred syscalls, \sys dynamically adjusts the selection probability of each syscall according to its effectiveness in directed fuzzing.
Third, \sys does not insert the syscalls randomly but determines their orders based on the dependencies of different syscalls.
Based on the above methods, \sys can generate semantically correct inputs that have close relevance to the target.

Below, we will give a detailed description of the methods in \sys.

\subsection{Lightweight and Fine-grained Distance Calculation}\label{subsec:gfuzz-distance}

When applying the existing DGF techniques in OS kernels like \gv, there are two main issues with their static distance calculation methods.
One is that the time overhead of distance calculation is unacceptably high.
The other is that the distance information is coarse-grained and imprecise.
Next, we first provide a depth analysis on the two challenges in detail and then propose our solutions.

\textbf{High Time Overhead.}
The time overhead of existing DGF in calculating static distance is hardly affordable for large systems like OS kernels.
There are two main state-of-the-art directed greybox fuzzers: \aflgo and Hawkeye. 
As only \aflgo is open-source, we can only test \aflgo and analyze its code.
Next, we mainly take \aflgo as the example.
For instance, \aflgo spends nearly 2 hours in compiling and instrumenting \texttt{cxxfilt}~\cite{nguyen2020binary}.
Compared to \texttt{cxxfilt}, \gv is an OS kernel, which has more code.
In our experiments, we find that \aflgo spends more than 16 hours in average in calculating the static distance for \gv.
Note that when testing a new target, the distance information needs to be re-calculated.
To address the high time overhead issue, we conduct an in-depth analysis about the method and code of existing DGF in calculating distance and find the following causes.

First, \aflgo calculates the distance between all code and the target with a traversal method.
Nevertheless, only parts of the code can lead to the target.
Thus, it is unnecessary to calculate distances for all code.
Second, both \aflgo and Hawkeye utilize the Dijkstra algorithm to find the shortest path between two nodes to further calculate their distance.
The Dijkstra algorithm is mainly used to find the shortest path for weighted graphs.
As both CG and CFG are unweighted graphs, it is unnecessary to use the Dijkstra algorithm.
Third, the implementation code of \aflgo has many issues.
For instance, each time \aflgo calculates a function level distance, it starts a Python process, incurring unnecessary Python startup and library initialization time cost repeatedly.
Moreover, the static distance information is instrumented into the target program.
That is, the high time overhead distance calculation is tightly coupled to the compilation process.
When testing other targets, in addition to re-calculating the static distance, \aflgo has to re-build the target program, causing much overhead.

\textbf{Coarse-grained and Imprecise Distance.}
Despite the time overhead issue, the distance information calculated by the existing DGF methods is inadequate in terms of granularity and precision.
First,  \aflgo and Hawkeye approximate basic block level distance with function level distance, which is coarse-grained and may cause much bias.
Second, when constructing CG, \aflgo does not consider the indirect calls.
Thus, the CG may miss some edges, which further affects the precision of the distance information.
Hawkeye uses \emph{Andersen}'s pointer analysis method~\cite{andersen1994program} to identify indirect calls.
However, pointer analysis is computational expensive~\cite{lu2019does} and does not guarantee the soundness~\cite{biallas2013ptrtracker, farkhani2018effectiveness}.

\textbf{Solutions of {\sc \textbf{G-Fuzz}}\xspace.}
To address the above issues, we propose a lightweight and fine-grained distance calculation method.
First, \sys constructs the CG of \gv and performs intra-procedural analysis to extract the CFG of each function in \gv.
We leverage \emph{Rapid Type Analysis (RTA)}~\cite{rta}, a type analysis method, and customize  \texttt{go-callvis}~\cite{go-callvis} to identify the indirect calls in \gv.
Compared to pointer analysis, type analysis is faster and has fewer false negatives.
We find that there exist false negatives (i.e., missing some critical indirect calls) when using pointer analysis for gVisor.
Missing critical indirect calls may make CG lose many edges, which will severely impact the guidance of directed fuzzing.
Thus, we choose to use type analysis although it may over-approximate some indirect calls.
Based on the extracted CG, \sys performs \emph{reachability analysis} on \gv for the given target to find out which functions are on the paths to the target.
We name these functions as \emph{reachable functions}.
In specific, starting from the target function, \sys adopts a bottom-up method to traverse the callers iteratively to obtain all the \emph{reachable functions}.
Then, based on the CG, \sys connects the CFGs for these  \emph{reachable functions} to construct a local inter-procedural CFG.
Next, \sys utilizes the \emph{BFS} algorithm to find the shortest path between each node of the local inter-procedural CFG and the target node. The length of the shortest path is the basic block level distance.

In summary, our proposed method can effectively address the issues of high time overhead and low precision in distance calculation.
For the time overhead issue, first, by performing \emph{reachability analysis}, we reduce the overhead caused by the unrelated code.
Second, we use the \emph{BFS} algorithm rather than the Dijkstra algorithm to find the shortest paths, which has a lower complexity.
More specifically, the complexity of the Dijkstra algorithm is $O(V^{2})$, while the complexity of the \emph{BFS} algorithm is $O(V+E)$, where $V$ represents the number of nodes in the graph, and $E$ represents the number of edges in the graph.
For the low precision issue, first, we utilize type analysis to identify the indirect calls of CG to reduce the false negatives.
Second, based on the \emph{reachability analysis}, we can effectively construct a local inter-procedural CFG to calculate the more precise basic block level distance than existing DGF.
In addition, 
we do not instrument the distance information into the \gv, but use a file to map each basic block and its static distance, making the method more scalable than existing DGF methods. 
We conduct experiments to provide detailed evaluation and analysis about the overhead and precision of our distance calculation method.
Specifically, for six different \gv's targets, \sys spends 97.1 seconds in calculation the static distance for \gv in average, while \aflgo spends 68346 seconds in average, demonstrating the significant performance of \sys.
More detailed results are presented in \S \ref{subsec:distance}.

\begin{figure}[!t]
    \lstset{frame=single,
        basicstyle=\scriptsize,
        language=Go,
        breaklines=true,
        numberstyle=\tiny,
        keywordstyle=\color{blue!90},
        commentstyle= \color{red!90}, 
        rulesepcolor= \color{ red!20!green!20!blue!20},
        postbreak=\mbox{\textcolor{red}{$\hookrightarrow$}\space},
    }
    \begin{lstlisting}[language=Go, numbers=left]
func (rfd *replicaFileDescription) Ioctl(         ctx context.Context, io usermem.IO,      args arch.SyscallArguments) (uintptr, error) {
    ...
    switch cmd := args[1].Uint(); cmd {
        case linux.FIONREAD: 
            return 0, rfd.inode.t.ld.inputQueueReadSize(t, io, args)
        case linux.TCGETS:
            return rfd.inode.t.ld.getTermios(t, args)
        case linux.TCSETS:
            return rfd.inode.t.ld.setTermios(t, args)
    ...
}
    \end{lstlisting}
    \caption{A code snippet related with \texttt{ioctl}.}
    \vspace*{-5mm}
    \label{fig-ioctl}
\end{figure}

\begin{table}[tbp]
\centering 
\footnotesize
\caption{The information of the syscall variants that \sys can infer.}\label{tab:extend}
\setlength{\tabcolsep}{10pt}{
\scalebox{0.9}{
\begin{tabular}{c|c|l}
\hline
\textbf{Syscall} & \begin{tabular}[c]{@{}c@{}}\textbf{Variants}\\ \textbf{Count}\end{tabular} & \multicolumn{1}{c}{\textbf{Example}}                                \\ \hline
arch\_prctl  & 3                    & arch\_prctl\$ARCH\_GET\_FS             \\ \hline
epoll\_ctl   & 3                    & epoll\_ctl\$EPOLL\_CTL\_MOD            \\ \hline
getsockopt   & 10                   & getsockopt\$inet6\_IPV6\_IPSEC\_POLICY \\ \hline
ioctl        & 49                   & ioctl\$TCGETS2                         \\ \hline
prctl        & 21                   & prctl\$PR\_SET\_MM\_AUXV               \\ \hline
semctl       & 15                   & semctl\$SEM\_INFO                      \\ \hline
setsockopt   & 16                   & setsockopt\$inet\_MCAST\_JOIN\_GROUP   \\ \hline
shmctl       & 9                    & shmctl\$SHM\_LOCK                      \\ \hline
waitid       & 1                    & waitid\$P\_PIDFD                       \\ \hline
\end{tabular}}}
\vspace{-2mm}
\end{table}

\subsection{Inference of Target Related Syscalls}

In essence, the goal of directed fuzzing \gv is to generate the syscall sequences that can trigger the specified target.
However, generating the inputs which can trigger the target from the whole input space is quite challenging.
\gv implements more than 200 Linux syscalls, and each syscall may have multiple parameters.
Moreover, a sequence may contain different numbers of syscalls, and the permutations of their order are various.
Thus, the whole input space is extremely vast.
To reduce the input space needed to explore and improve the efficiency of directed fuzzing, we propose an automatic method to infer which syscalls are related to the given target before fuzzing.
This method is based on static analysis and the expert knowledge of \gv. 
{\color{black}  
It needs to note that the generation process of the inference rules relies on the expert knowledge.
But once the inference rules are formed, they can be used automatically.
The usage process is as follows: a user just provides the specific location of the target or crash report\footnote{For bug reproduction, if a user has the crash report, the eighth inference 
 rule \emph{stack trace based syscall inference} will be used.}, and the inference rules automatically output its related syscalls.
During the next dynamic fuzzing process, the utilization of the inferred syscalls is also automatic.
}
\sys has incorporated the following eight inference rules and can be further extended to incorporate more rules.

\textbf{Function Call Chain based Inference.}
This is a general rule for all the targets of \gv.
First, based on the \emph{reachability analysis} of the given target, we can obtain the \emph{reachable functions} that lead to the target.
Some of the reachable functions are syscall handlers, and the corresponding syscalls may be related to the target.
According to the information of files \texttt{pkg/sentry/syscalls/linux/linux64.go} and \texttt{pkg/sentry/syscalls/linux/vfs2/vfs2.go} in \gv code, we construct the mapping table of the functions and the corresponding syscalls.
Second, we extract the related syscalls from the \emph{reachable functions}.
Note that not all these syscalls have close relevance to the target.
To reduce false positives, in this rule, we only select the syscalls of the function that has the smallest distance to the target as the inference results.

\textbf{Specialized Syscall Inference.}
The syscalls may have multiple parameters, and some of them may have complex structures and various values.
To provide precise and semantically correct mutation, Syzkaller provides the syscall variants~\cite{syscal-des} by instantiating partial arguments of the original syscalls.
For instance,  \texttt{ioctl\$FIONREAD} is a variant of syscall \texttt{ioctl}, which denotes the value of its parameter \texttt{request} is set as \texttt{FIONREAD} (the concrete value is 1074030207).
When Syzkaller mutates this variant, the \texttt{request} parameter will not be modified.
For original syscall \texttt{ioctl}, Syzkaller implements 411 syscall variants for it.
If we successfully infer that the \texttt{ioctl} is related to a target but not be able to infer the concrete parameters, it is still challenging to reach this target.

In total, Syzkaller implements 948 syscall variants for 47 original syscalls.
Thus, to reduce the input space, we have to infer the target related syscall variants.
By conducting an in-depth analysis of the \gv code, we find that the variant information can be extracted from the related code paths.
Fig.~\ref{fig-ioctl} illustrates an example code that is related with \texttt{ioctl}.
From the constants of the \texttt{case} statements, we can extract the concrete variants of \texttt{ioctl}.
For instance, the code on line 5 is related with syscall variant \texttt{ioctl\$FIONREAD}, and the code on line 7 is related with syscall variant \texttt{ioctl\$TCGETS}.
Thus, we infer the syscall variants by extracting the special constants from the code path towards the target.
Notably, in the inference process, if an inferred syscall has variants, we then perform further inference about its variants, and add the syscall variants into the inference results.
Based on this method, we can infer 127 variants for nine syscalls accurately.
Table~\ref{tab:extend} presents the information about the nine syscalls.

\begin{figure}[tbp]
    \centering
    \includegraphics[width=3.5in]{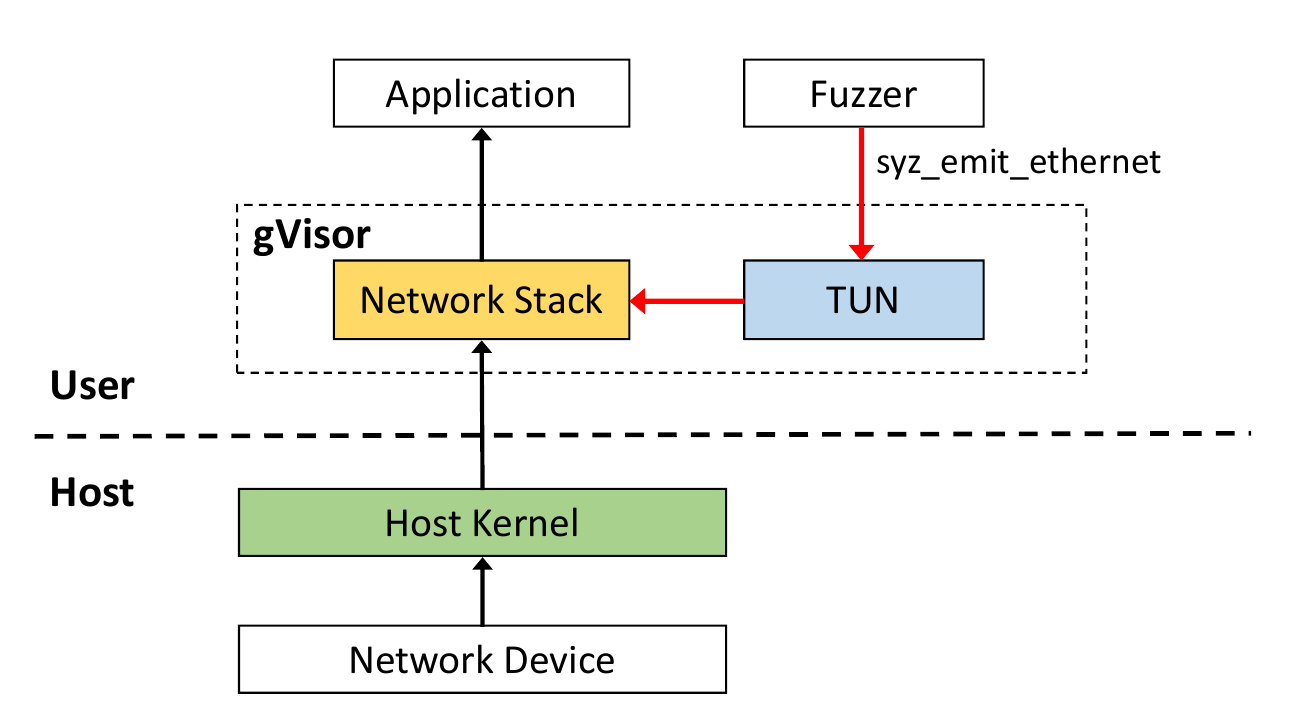}
    \caption{The workflow of \gv in processing network packets.}
    \vspace{-2mm}
    \label{fig:network}
\end{figure}

\lstdefinelanguage{syz}{
    morekeywords={syz_emit_ethernet, eth_packet, eth_payload, eth2_packet, ipv4_packet},
    sensitive=false, 
    morecomment=[l]{}
} %

\begin{figure}[!t]
    \lstset{frame=single,
        basicstyle=\scriptsize,
        language=C,
        breaklines=true,
        numberstyle=\tiny,
        keywordstyle=\color{blue!90},
        commentstyle= \color{red!90}, 
        rulesepcolor= \color{ red!20!green!20!blue!20},
        linewidth=8.8cm
    }
    \begin{lstlisting}[language=syz, numbers=right]
syz_emit_ethernet(
  len    len[packet], 
  packet ptr[in, eth_packet], 
  frags  ptr[in, vnet_fragmentation, opt])

eth_packet {
  dst_mac  mac_addr
  src_mac  mac_addr
  vtag     optional[vlan_tag]
  payload  eth_payload
} [packed]

eth_payload {
  eth2     eth2_packet
} [packed]

eth2_packet [
  generic  eth2_packet_generic
  arp      eth2_packet_t[ETH_P_ARP, arp_packet]
  ipv4     eth2_packet_t[ETH_P_IP, ipv4_packet]
  ipv6     eth2_packet_t[ETH_P_IPV6, ipv6_packet]
  llc      eth2_packet_t[ETH_P_802_2, llc_packet]
  ... // 7 more union selections omitted
] [varlen]
    
ipv4_packet [
  generic ipv4_packet_t[flags[ipv4_types,int8],                   array[int8]]
  tcp     ipv4_packet_t[const[IPPROTO_TCP,int8],                   tcp_packet]
  udp     ipv4_packet_t[const[IPPROTO_UDP,int8],                   udp_packet]
  icmp    ipv4_packet_t[const[IPPROTO_ICMP,int8],                   icmp_packet]
  ... // 4 more union selections omitted
] [varlen]
    \end{lstlisting}
    \caption{The parameters of {syz\_emit\_ethernet}.}
    \vspace{-2mm}
    \label{fig-sysemit}
\end{figure}

\textbf{Network Packet Processing Related Syscalls.}
\gv implements a user-space network stack (i.e., \texttt{netstack}) to process the network packets between the host kernel and the container applications.
Fig.~\ref{fig:network} illustrates the how \gv processes the network packets.
In the practical scenario, the host kernel receives the network packets from the network device, and then transfers the packets to the \texttt{netstack} of \gv. 
In essence, the inputs to trigger of \texttt{netstack} come from the network devices, not the upper syscalls.
Nevertheless, deploying the real network devices to generate inputs to trigger the targets of \texttt{netstack} is not practical.
To solve this problem, Syzkaller designs an extra syscall named \texttt{syz\_emit\_ethernet}.
As red arrows in Fig.~\ref{fig:network} shows, \texttt{syz\_emit\_ethernet} leverages the \texttt{TUN}, a virtual network device to directly inject network packets to \texttt{netstack}.
Thus, we can leverage the \texttt{syz\_emit\_ethernet} syscall to test the targets of \gv network stack code.

Nevertheless, the parameters of \texttt{syz\_emit\_ethernet} syscall are complex, which are related with many nested \texttt{unions} and \texttt{structs}. 
As Fig.~\ref{fig-sysemit} shows, there are three parameters of \texttt{syz\_emit\_ethernet}.
Taking the second parameter \texttt{packet ptr} as example,
packet \texttt{ptr} points to a \texttt{struct} type data structure, which consists of four members.
The fourth member variable \texttt{eth\_payload} is a \texttt{struct}, and its member \texttt{eth2\_packet} is an \texttt{union}.
\texttt{eth2\_packet} includes 12 members, and its third memeber \texttt{ipv4} is also an \texttt{union}.
\texttt{ipv4\_packet}, a member of \texttt{ipv4} is also an \texttt{union}, which consists of eight members.
As a consequence, the parameters of \texttt{syz\_emit\_ethernet} are nested and complex.
Only successfully inferring the target related syscall \texttt{syz\_emit\_ethernet} is not enough nor efficient. 
The whole input space of \texttt{syz\_emit\_ethernet} is vast due to the complex parameters.

To address this problem, we conduct in-depth analysis on the \texttt{syz\_emit\_ethernet} related targets.
We find that some parameters of \texttt{syz\_emit\_ethernet} can be inferred based on the information of the targets.
For instance, for the target function \texttt{handleICMP} that locates in the directory \texttt{network/ipv4/icmp.go} of \gv, intuitively, it is related with \texttt{ipv4} and \texttt{icmp}.
Moreover, some parameters are mutually exclusive from each other.
For instance, if the target is related with \texttt{ipv4}, it may have less relevance to \texttt{ipv6}.
For the target function \texttt{handleICMP}, we can set the parameter \texttt{eth2\_packet} as \texttt{ipv4}, and set the parameter \texttt{ipv4\_packet} as \texttt{icmp\_packet}, with high probability.
In this way, we can reduce the input space efficiently.
According to the different parameters, we implement ten syscall variants for \texttt{syz\_emit\_ethernet}.
The detailed information of the ten syscall variants is shown in Table~\ref{tab:sysemit}.

\begin{table}[!t]
\centering 
\footnotesize
\caption{The variants of syscall \texttt{syz\_emit\_ethernet} in G-Fuzz.}\label{tab:sysemit}
\setlength{\tabcolsep}{8pt}{
\scalebox{0.88}{
\begin{tabular}{l|l}
\hline
\multicolumn{1}{c|}{\textbf{Variants}}    & \multicolumn{1}{c}{\textbf{Union Selections}}        \\
\hline
syz\_emit\_ethernet\$ipv4       & eth2\_packet=ipv4           \\
syz\_emit\_ethernet\$ipv4\_tcp  & eth2\_packet=ipv4 \&\& ipv4\_packet=tcp  \\
syz\_emit\_ethernet\$ipv4\_udp  & eth2\_packet=ipv4 \&\& ipv4\_packet=udp  \\
syz\_emit\_ethernet\$ipv4\_icmp & eth2\_packet=ipv4 \&\& ipv4\_packet=icmp \\
syz\_emit\_ethernet\$ipv4\_igmp & eth2\_packet=ipv4 \&\& ipv4\_packet=igmp \\
syz\_emit\_ethernet\$ipv6       & eth2\_packet=ipv6                        \\
syz\_emit\_ethernet\$ipv6\_tcp  & eth2\_packet=ipv6 \&\& ipv6\_packet=tcp  \\
syz\_emit\_ethernet\$ipv6\_udp  & eth2\_packet=ipv6 \&\& ipv6\_packet=udp  \\
syz\_emit\_ethernet\$ipv6\_icmp & eth2\_packet=ipv6 \&\& ipv6\_packet=icmp \\
syz\_emit\_ethernet\$arp        & eth2\_packet=arp          \\
\hline
\end{tabular}}}
\end{table}

\textbf{Virtual File System Related Syscalls.}
\gv and other Unix-like kernels implement virtual file systems such as \texttt{proc}, \texttt{devpts}, etc.
Different from the file systems such as \texttt{ext4}, \texttt{FAT32}, virtual file systems are virtual and do not locate in the disk space.
The content of the virtual file systems resides in memory.
The container applications use the interfaces such as \texttt{open}, \texttt{read}, \texttt{write} to communicate with the kernel.
Specifically, by setting different parameters, applications can request different file operations.
If the targets reside in the virtual file system implementation code of \gv, the corresponding syscalls to trigger them are the common interfaces.

Nevertheless, only knowing the interfaces is not enough to trigger these targets.
These common interfaces can be invoked by much code.
For instance, taking the function
\texttt{masterFileDescription.Read} of \texttt{devpts} virtual file system as the target, the syscall to trigger this target is \texttt{read} and the its first parameter should be a \texttt{devpts} type file descriptor.
We can use syscall \texttt{openat\$ptmx} to generate a \texttt{devpts} related file descriptor.
Thus, in addition, to infer the \texttt{read} syscall, we have to infer the corresponding parameter type.
Based on the expert knowledge, we provide the inference rules for six types of the most used virtual file systems, which are shown in Table~\ref{tab:gfuzz-vfilesys}.

\begin{table}[tbp]
\centering 
\footnotesize
\caption{Different Virtual File System Related syscalls.}\label{tab:gfuzz-vfilesys}
\setlength{\tabcolsep}{10pt}{
\scalebox{0.95}{
\begin{tabular}{c|c}
\hline
\multicolumn{1}{c|}{\textbf{\begin{tabular}[c]{@{}c@{}}Virtual\\ Filesystem\end{tabular}}} & \multicolumn{1}{c}{\textbf{Inferred Syscalls}}                                   \\ \hline
devpts     & openat\$ptmx, syz\_open\_pts                        \\ \hline
eventfd    & eventfd2, eventfd                                   \\ \hline
kernfs     & syz\_open\_procfs                                   \\ \hline
pipefs     & pipe, pipe2                                         \\ \hline
signalfd   & signalfd4, signalfd                                 \\ \hline
timerfd    & timerfd\_create, timerfd\_settime, timerfd\_gettime \\ \hline
\end{tabular}}}
\end{table}

\textbf{Error Handling Code Related Syscalls.}
There is much error handling code in OS kernels including \gv, which is critical and may contain serious security problems~\cite{jiang2020fuzzing, wu2021understanding}.
It is vital to test error handling code.
However, the current mechanism of Syzkaller cannot effectively test the error handling code.
For instance, Fig.~\ref{fig-copy} presents an example of memory copy error handling code in \gv. 
The function \texttt{primitive.CopyInt32SliceOut} at line 3 is to copy the third parameter \texttt{fds} to the second parameter \texttt{addr}.
{color{red}
To trigger the target that resides in the code of line 4 to 9, the input sequence must make function \texttt{primitive.CopyInt32SliceOut} eturn an error.}
Syzkaller generates the concrete value of the syscall parameter based on its type.
For this instance, the parameter \texttt{addr} is a pointer type, and Syzkaller tends to assign a valid memory address for \texttt{addr}.
However, this target branch can not be triggered with a valid memory address.

To trigger the error handling code, intuitively, it should increase the probability of ``causing an error".
Here, we mainly focus on two common types of error handling code of \gv: 1) memory related and 2) permission check related error handling code.
For memory related error handling code, we can add three syscalls: \texttt{mmap}, \texttt{munmap} and \texttt{mprotect} into the inferred results to make some memory invalid.
Specifically, setting the parameter \texttt{prot} of syscalls \texttt{mmap} and \texttt{mprotect}  as \texttt{PROT\_NONE}, or using syscall \texttt{munmap} to release memory,  can likely make the pointer type parameter points to the invalid memory.

For permission check related error handling code, we can add two syscalls: \texttt{setuid} and \texttt{setresuid} to change the current user of the running process.
Thus, the syscalls after \texttt{setuid} or \texttt{setresuid} are more likely to have no access to the previously created resources, which increases the probability of permission check errors.

\begin{figure}[!t]
    \lstset{frame=single,
        basicstyle=\scriptsize,
        language=Go,
        breaklines=true,
        numberstyle=\tiny,
        keywordstyle=\color{blue!90},
        commentstyle= \color{black!90}, 
        rulesepcolor= \color{ red!20!green!20!blue!20},
    }
    \begin{lstlisting}[language=Go, numbers=right]
func pipe2(t *kernel.Task, addr usermem.Addr, flags int32) error {
    ...
    if _, err := primitive.CopyInt32SliceOut(t, addr, fds); err != nil {
        for _, fd := range fds {
            if _, file := t.FDTable().Remove(t, fd); file != nil {
                file.DecRef(t)
            }
        }
        return err
    }
    return nil
}
    \end{lstlisting}
    \caption{An example of memory copy error handling code in \gv.}
    \vspace{-2mm}
    \label{fig-copy}
\end{figure}

\textbf{Seccomp Related Syscalls.}
\emph{Seccomp} (Secure computing mode) is a security mechanism in Linux kernel, which is also adopted in \gv. 
By setting the filter rules, \emph{seccomp} restricts the syscalls that are allowed to be accessed by the current process.
In particular, programmers can specify which system  calls are permitted by writing \emph{Berkeley Packet Filter (BPF)} programs.
The \emph{BPF} related code in \gv resides in directory \emph{pkg/bpf}, which is to parse and execute the \emph{BPF} rules.
However, these code are not directly executed when registering \emph{BPF} rules, but are triggered by a hook mechanism afterward.
Before subsequent syscalls of the current process can be executed, the hook will first execute all registered \emph{BPF} rules to determine whether the syscall is allowed, which will trigger the code under the directory \emph{pkg/bpf}.
Thus, the \emph{Function Call Chain based} inference rule cannot effectively find these registering syscalls.
To tackle this problem, when testing this code as the target, we add the four \emph{BPF} registering syscalls as the inference results, which are \texttt{prctl\$PR\_SET\_SECCOMP},
\texttt{seccomp\$SECCOMP\_SET\_MODE\_STRICT},
\texttt{seccomp\$SECCOMP\_SET\_MODE\_FILTER} and
\texttt{seccomp\$SECCOMP\_SET\_MODE\_FILTER\_LISTENER}. 
These syscalls can register the \emph{BPF} hook for subsequent syscalls or change the operating mode of seccomp of the current process.
It is worth noting that there are more syscalls that are related to \emph{BPF} mechanism, like \texttt{bpf}.
Nevertheless, \gv does not support these syscalls by directly returning capability errors.
Therefore, we do not include them in the inference results.

 \textbf{Readiness Mechanism Code Related Syscalls.}
In order to achieve high-performance I/O operations, the Linux kernel proposes \emph{epoll} mechanism to provide scalable I/O event notification.
The applications do not need to repeatedly query the status of the file descriptor to see whether I/O operation is possible, but instead, wait for the kernel to notify the state changes of the file descriptors.
\gv also implements this \emph{epoll} mechanism, and adds a virtual file system level interface named \texttt{vfs.FileDescriptionImpl.Readiness}.
All virtual file systems implement this interface to return whether the current file is readable or writable.
With the understanding of this epoll mechanism, for Readiness function-related targets, 
that is, the target itself is implementing \texttt{vfs.FileDescriptionImpl.Readiness} interface or called by a Readiness function, 
we should add epoll-related syscalls into our inference results, 
including \texttt{pselect6}, \texttt{epoll\_ctl\$EPOLL\_CTL\_ADD}, \texttt{ppoll} and \texttt{poll}.

\begin{figure}[!t]
\vspace{-2mm}
\setlength{\belowcaptionskip}{-1mm}
    \centering
    \includegraphics[width=3.3in]{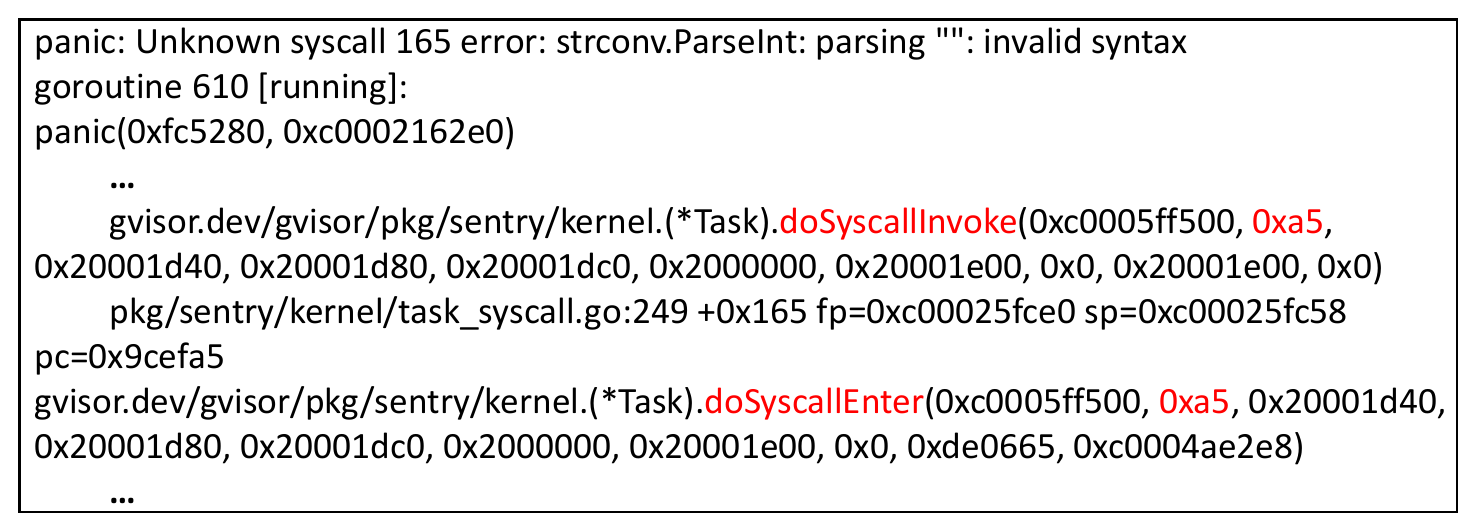}
    \caption{The stack trace information of a \gv bug.}
    \vspace{-3mm}
    \label{fig:crash-rep}
\end{figure}

\textbf{Stack Trace Based Syscall Inference.}
Bug reproduction is a typical application scenario of directed fuzzing.
Compared to the general directed fuzzing scenarios that are only provided with the target locations, 
bug reproduction scenarios usually have more information, such as the stack trace information when the bug is triggered.
Thus, we should make full use of this information when reproducing the bugs.
Fig.~\ref{fig:crash-rep} illustrates the stack trace and a PoC information of a \gv bug.
We can extract the bug related syscalls from its stack trace.
For instance, the second parameter (i.e., 0xa5) of function \texttt{doSyscallInvoke} or \texttt{doSyscallEnter} represents the
ID of syscall \texttt{mount}, and the PoC of this bug contains the syscall \texttt{mount}.
Intuitively, for this bug, we can reproduce it effectively if we can infer its related syscalls.

Motivated by the above instance, for the target of bug reproduction, we infer its related syscalls from the stack trace (if given).
First, similar to the function call chain based inference method, we extract the related syscalls from the functions of the stack trace.
Second, we extract the related syscalls from the arguments of special functions such as \texttt{doSyscallInvoke} to find the corresponding syscalls.
By incorporating the extracted syscalls into the inference results, \sys leverages the information from the stack trace to facilitate bug reproduction.

\subsection{The Utilization of the Inferred Syscalls}\label{subsec:utilization}

In directed fuzzing process, the inferred syscalls are utilized in the mutation process to generate new inputs.
We define the probability that whether we choose to use the inferred syscalls in the mutation as $p$.
The value of $p$ decreases linearly during the fuzzing process.
Assume the maximum value of $p$ is $p_{max}$, and the minimum value of $p$ is $p_{min}$.
The fuzzing timeout is $T_{fuzz}$.
The value of $p$ at the current fuzzing time $t$ is:

\begin{equation}\label{equ-useinfer}
p = p_{max} - \frac{p_{max}-p_{min}}{T_{fuzz}}\ast t
\end{equation}

In our experiment, we set $p_{max}=0.9$ and $p_{min}=0.1$.
At the beginning of directed fuzzing, the probability of utilizing the inferred syscalls is 90\%.
With the fuzzing time increases, the probability decreases linearly until 10\%.
The rationale behind this design choice is that if we cannot trigger the target in a relatively long time with the inferred syscalls, 
the inference results may have false positives and further mislead the fuzzing process.
Thus, we should decrease the probability of using the inferred syscalls.

The inferred syscalls are used to insert into a selected seed to generate the new inputs.
To make efficient use of the inferred syscalls and generate semantically correct inputs, we propose two methods in the mutation stage: \emph{Selection Probability Scheduling Strategy} and \emph{Insert Order Scheduling Strategy}.

 \textbf{Selection Probability Scheduling Strategy.}
There may exist false positives in the inference results.
In other words, some inferred syscalls may not be related to the target in fact, which may reduce the efficiency of directed fuzzing.
To alleviate the impacts of the false positives, we propose a strategy to dynamically adjust the selection probability of each inferred syscall in the mutation process.

Assume the set of the inferred syscalls for the given target is $S=\{s_{1}, s_{2},...,s_{n}\}$, where $n$ is the number of all inferred syscalls.
The initial weight $W$ of each inferred syscall is one, which is equal.
That is, $W_{S_1}=W_{S_2}=...=W_{S_n}=1$.

During the fuzzing process, the inputs which can trigger new paths will be saved in a \emph{global seed queue}.
For the new saved input, we will determine whether its distance is shorter than its parent seed.
If so, we will save it in another seed queue named \emph{shorter distance queue}.
Intuitively, the seed in \emph{shorter distance queue} may contain the syscalls that can be helpful to reach the target.
For each inferred syscall, we count its frequency in \emph{shorter distance queue}.
An incorrect inferred syscall is highly unlikely to appear in this queue.
Assume at fuzzing time $t$, the frequency of an inferred syscall $s_{i}$ is $F_{(s_{i}, t)}$.
At time $t$, the inferred syscall $s_{i}$'s weight $W_{s_{i}}(t)$ is $F_{(s_{i}, t)}+1$.
The selection probability of the syscall $s_{i}$ is calculated by formula~\ref{equ-prob}.

\begin{equation}\label{equ-prob}
    P_{s_i}(t)=\frac{W_{s_i}(t)}{\sum_{j=1}^{n}W_{s_j}(t)} 
\end{equation}

As the fuzzing goes on,  the probability of selecting the correct inferred syscalls increases, and the  probability of selecting the incorrect inferred syscalls decreases.
In this way, \sys can effectively mitigate the negative impacts of incorrect inference results.

 \textbf{Insert Order Scheduling Strategy.}
The order of the syscalls has a significant impact on the semantic of the whole input.
There may exist dependencies between different syscalls.
MoonShine~\cite{pailoor2018moonshine} classifies the dependency relationships between different syscalls into two types: \emph{explicit} and \emph{implicit} dependencies.
The explicit dependencies represent that the return value is used as an argument of another syscall.
The implicit dependencies represent that there are shared data structures among syscalls.
Thus, to generate the semantically correct inputs as possible, the syscalls dependent on others should be placed at the front position of the entire sequence.

MoonShine focuses on generating the initial seeds with offline static analysis.
In contrast, we aim to generate inputs by mutating seeds during the fuzzing process, which is an online and time-sensitive task.
To provide fast mutation,  
we mainly consider the \emph{explicit} dependencies.
We regard the syscalls that generate the dependent return values as the ``producers" and regard the syscalls that receive the dependent values as the ``consumers".
Intuitively, the ``producer" syscalls should be inserted in front of the ``consumer" syscalls.
To this end, we use the function \texttt{biasedRand} provided by Syzkaller to generate the insert index for ``consumer" syscalls.
In specific, \texttt{biasedRand(n, k)} is to randomly generate a number, which conforms that the probability of outputting n-1 is k times higher than the probability of outputting 0.
We set k as 5 in the experiments.
For the ``producer" syscall, we use \texttt{n - biasedRand(n, k)} to generate its insert index.
In this way, the ``producer" syscalls tend to be inserted in front of the ``consumer" syscalls.

\subsection{Exploration and Exploitation Dynamic Switch}\label{sub:eeswitch}

The core idea of directed greybox fuzzing is using distance information to guide the fuzzer to constantly generate test cases closer to the target until triggering it.
However, distance information has limitations in guiding directed fuzzing.
As we discussed in \S \ref{subsec:gfuzz-distance}, distance information has low precision.
Despite the issues caused by distance calculation methods, existing static analysis technique cannot identify the indirect calls with no error~\cite{lu2019does}. 
In addition, distance cannot be able to measure the difficulty to reach the target. 
Thus, it is necessary to make reasonable use of the distance information.

One practical solution is to incorporate the \emph{exploitation} with \emph{exploration}.
Here the \emph{exploitation} represents the typical ``directed fuzzing", which focuses on testing the shortest or closer paths to the target.
The \emph{exploration} represents the typical ``coverage-based fuzzing", which focuses on exploring the new paths.
The \emph{exploration} can alleviate the errors of the distance and can avoid directed fuzzing trapping into the local optimal.
Nevertheless, overusing \emph{exploration} makes directed fuzzing degenerate into the coverage-based fuzzing, which is not efficient.

Thus, how to coordinate the \emph{exploitation} and \emph{exploration} to amplify their advantages in directed fuzzing is critical.
AFLGo~\cite{bohme2017directed} adopts a fixed time-wise splitting method to balance the exploration and exploitation phases.
Specifically, for a 24-hour fuzzing experiment, \aflgo sets the exploration time as 20 hours and the exploitation time as 4 hours.
However, this fixed setting may not suit all the conditions in directed fuzzing, such as testing different programs or targets.
Paper~\cite{ye2020rdfuzz} shows that the performance of \aflgo varied much with different settings in exploration time.
Therefore, the solution of \aflgo is not satisfying.
Other fuzzing work such as EcoFuzz~\cite{yue2020ecofuzz} also studies the \emph{exploitation} and \emph{exploration} trade-off problem, whereas it focuses on solving the seed power scheduling problem, not the directed fuzzing.

To tackle this problem, we propose a \emph{exploration and exploitation dynamic switch} method, which coordinates the two phases adaptively according to the current state of fuzzing.
When one phase does not make any contribution to fuzzing for a time threshold, \gv will switch to the other phase, and vice versa.
The detailed workflows of this method are the following.

At the start of directed fuzzing, \sys first enters the initial phase, generating and executing the initial inputs.
The inputs that can discover new paths will be saved in the seed pool.
After this phase, \sys first enters the closer seed exploitation phase.
In this phase, \sys randomly selects $m$ seeds from the seed pool and selects the top $k$ seeds with the shortest distance.
Note that the distance of a seed is the shortest distance of its executed basic blocks.
The shortest $k$ seeds are mutated to generate new inputs.
Then, \sys executes the new inputs. 
For the inputs that trigger the new paths, \sys saves them in the seed pool.
During the fuzzing process, \sys monitors whether the current phase gets into a ``stuck" state.
If so, \sys will switch into the other phase.
For the closer seed exploitation phase, the ``stuck" state is not finding any new paths of the reachable set over the time threshold $T_{a}$.
When \sys switches into the path exploration phase, it does not consider the distance of each seed.
In specific, it selects seeds from the seed pool according to the coverage of each seed.
The seed with more coverage has a higher probability of being selected.
During this phase, if a seed finds new paths, before saving it in the seed pool, it will be directly mutated to generate the new inputs.
The ``stuck" state of this phase is not finding any new paths in time threshold $T_{b}$.
Where $T_{a}$ and $T_{b}$ are user-defined.
In our experiments, we set $T_{a}$ to five minutes, and $T_{b}$ to ten minutes.

\section{Implementation of  \sys} 

\sys contains two components: static analysis and directed fuzzing.
The functionalities of the static analysis include CG/CFG extraction, reachability analysis, distance calculation and target related syscalls inference.
We implement the CG extraction based on \texttt{go-callvis}~\cite{go-callvis}.
For the CFG of \gv, we first use \texttt{GoLLVM}~\cite{gollvm} to compile \gv into LLVM IR, an intermediate representation.
Then, we implement an LLVM Pass to construct the target related inter-procedural CFG. The Pass has 1,238 lines of C++ code.
The distance calculation of \sys is implemented with 313 lines of Python code.
The target related syscalls inference is written with 212 lines of Python code.
The directed fuzzing component of \sys is implemented based on the state-of-the-art OS kernel fuzzer  Syzkaller (commit ID=9d751681c).
We add more than 1,500 lines of Go language code, comparing with the origin Syzkaller.

\section{Evaluation}\label{sec:evaluation}

In this section, we mainly evaluate the performance of \sys in directed fuzzing for \gv.

\textbf{Basic Settings.}
As there exists randomness in fuzzing experiments, following the guidance in paper~\cite{unifuzz-li}, we conduct each experiment with 20 repetitions.
For general target testing and patch testing, the fuzzing time is 24 hours.
For bug reproduction, as it is more difficult, the fuzzing time is 72 hours.

\begin{table}[!h]
\setlength{\belowcaptionskip}{0cm}
\centering 
\footnotesize
\caption{The performance of \sys, Syz-Go and Syzkaller in general target testing.}\label{tab:normal-target}	
\setlength{\tabcolsep}{6pt}{
\scalebox{0.9}{
			\begin{tabular}{clccccc}
				\hline
\textbf{Target   ID} & \textbf{Fuzzer} & \textbf{runs} & \textbf{$\mu$TTE (h)} & \textbf{Speedup}    & \textbf{$\hat{A}_{12}$}  & \textbf{p-value} \\ \hline
\multirow{3}{*}{\#1} & G-Fuzz     & \textbf{18} & \textbf{6.79} & - & - & -\\ \cline{2-7}
 & Syz-Go & \textbf{18} & 12.43 & 1.83 & 0.79 & \textless{}0.01\\ \cline{2-7}
 & Syzkaller & 12 & 17.26 & 2.54 & 0.87 & \textless{}0.01\\ \specialrule{0.5pt}{1pt}{1pt}
\multirow{3}{*}{\#2} & G-Fuzz     & \textbf{20} & \textbf{0.92} & - & - & -\\ \cline{2-7}
 & Syz-Go & 16 & 15.06 & 16.39 & 0.99 & \textless{}0.01\\ \cline{2-7}
 & Syzkaller & 7 & 20.57 & 22.38 & 1.00 & \textless{}0.01\\ \specialrule{0.5pt}{1pt}{1pt}
\multirow{3}{*}{\#3} & G-Fuzz     & \textbf{20} & \textbf{0.15} & - & - & -\\ \cline{2-7}
 & Syz-Go & 18 & 14.23 & 97.46 & 1.00 & \textless{}0.01\\ \cline{2-7}
 & Syzkaller & 8 & 19.27 & 131.96 & 1.00 & \textless{}0.01\\ \specialrule{0.5pt}{1pt}{1pt}
\multirow{3}{*}{\#4} & G-Fuzz     & 19 & \textbf{1.36} & - & - & -\\ \cline{2-7}
 & Syz-Go & \textbf{20} & 5.40 & 3.96 & 0.95 & \textless{}0.01\\ \cline{2-7}
 & Syzkaller & 14 & 15.42 & 11.31 & 0.96 & \textless{}0.01\\ \specialrule{0.5pt}{1pt}{1pt}
\multirow{3}{*}{\#5} & G-Fuzz     & \textbf{20} & \textbf{0.20} & - & - & -\\ \cline{2-7}
 & Syz-Go & \textbf{20} & 5.40 & 26.99 & 0.99 & \textless{}0.01\\ \cline{2-7}
 & Syzkaller & 17 & 11.55 & 57.80 & 1.00 & \textless{}0.01\\ \specialrule{0.5pt}{1pt}{1pt}
\multirow{3}{*}{\#6} & G-Fuzz     & \textbf{20} & \textbf{0.18} & - & - & -\\ \cline{2-7}
 & Syz-Go & 19 & 6.05 & 33.84 & 1.00 & \textless{}0.01\\ \cline{2-7}
 & Syzkaller & \textbf{20} & 8.23 & 46.09 & 1.00 & \textless{}0.01\\ \specialrule{0.5pt}{1pt}{1pt}
\multirow{3}{*}{\#7} & G-Fuzz     & \textbf{9} & \textbf{18.02} & - & - & -\\ \cline{2-7}
 & Syz-Go & 3 & 21.21 & 1.18 & 0.63 & 0.044\\ \cline{2-7}
 & Syzkaller & 5 & 21.28 & 1.18 & 0.61 & 0.077\\ \specialrule{0.5pt}{1pt}{1pt}
\multirow{3}{*}{\#8} & G-Fuzz     & \textbf{19} & \textbf{4.69} & - & - & -\\ \cline{2-7}
 & Syz-Go & 4 & 20.28 & 4.33 & 0.91 & \textless{}0.01\\ \cline{2-7}
 & Syzkaller & 2 & 22.91 & 4.89 & 0.95 & \textless{}0.01\\ \specialrule{0.5pt}{1pt}{1pt}
\multirow{3}{*}{\#9} & G-Fuzz     & \textbf{10} & \textbf{14.49} & - & - & -\\ \cline{2-7}
 & Syz-Go & 2 & 23.28 & 1.61 & 0.72 & \textless{}0.01\\ \cline{2-7}
 & Syzkaller & 0 & 24.00 & 1.66 & 0.75 & \textless{}0.01\\ \specialrule{0.5pt}{1pt}{1pt}
\multirow{3}{*}{\#10} & G-Fuzz     & \textbf{19} & \textbf{4.28} & - & - & -\\ \cline{2-7}
 & Syz-Go & 10 & 17.85 & 4.17 & 0.91 & \textless{}0.01\\ \cline{2-7}
 & Syzkaller & 6 & 20.30 & 4.74 & 0.93 & \textless{}0.01\\ \specialrule{0.5pt}{1pt}{1pt}
\multirow{3}{*}{\#11} & G-Fuzz     & \textbf{12} & \textbf{15.20} & - & - & -\\ \cline{2-7}
 & Syz-Go & 8 & 20.78 & 1.37 & 0.67 & 0.027\\ \cline{2-7}
 & Syzkaller & 5 & 19.92 & 1.31 & 0.66 & 0.032\\ \specialrule{0.5pt}{1pt}{1pt}
\multirow{3}{*}{\#12} & G-Fuzz     & \textbf{13} & \textbf{11.20} & - & - & -\\ \cline{2-7}
 & Syz-Go & 7 & 19.85 & 1.77 & 0.72 & \textless{}0.01\\ \cline{2-7}
 & Syzkaller & 5 & 21.98 & 1.96 & 0.78 & \textless{}0.01\\ \specialrule{0.5pt}{1pt}{1pt}
\multirow{3}{*}{\#13} & G-Fuzz     & \textbf{10} & \textbf{13.78} & - & - & -\\ \cline{2-7}
 & Syz-Go & 1 & 23.01 & 1.67 & 0.73 & \textless{}0.01\\ \cline{2-7}
 & Syzkaller & 0 & 24.00 & 1.74 & 0.75 & \textless{}0.01\\ \specialrule{0.5pt}{1pt}{1pt}
\multirow{3}{*}{\#14} & G-Fuzz     & \textbf{16} & \textbf{9.36} & - & - & -\\ \cline{2-7}
 & Syz-Go & 12 & 15.64 & 1.67 & 0.71 & 0.012\\ \cline{2-7}
 & Syzkaller & 12 & 14.43 & 1.54 & 0.61 & 0.111\\ \specialrule{0.5pt}{1pt}{1pt}
\multirow{3}{*}{\#15} & G-Fuzz     & \textbf{20} & \textbf{2.26} & - & - & -\\ \cline{2-7}
 & Syz-Go & 15 & 11.68 & 5.16 & 0.89 & \textless{}0.01\\ \cline{2-7}
 & Syzkaller & 15 & 11.61 & 5.13 & 0.91 & \textless{}0.01\\ \specialrule{0.5pt}{1pt}{1pt}
\multirow{3}{*}{\#16} & G-Fuzz     & \textbf{19} & \textbf{1.57} & - & - & -\\ \cline{2-7}
 & Syz-Go & 6 & 19.63 & 12.53 & 0.97 & \textless{}0.01\\ \cline{2-7}
 & Syzkaller & 9 & 18.21 & 11.63 & 0.96 & \textless{}0.01\\ \specialrule{0.5pt}{1pt}{1pt}
\multirow{3}{*}{\#17} & G-Fuzz     & \textbf{20} & \textbf{1.46} & - & - & -\\ \cline{2-7}
 & Syz-Go & 16 & 14.86 & 10.15 & 0.99 & \textless{}0.01\\ \cline{2-7}
 & Syzkaller & 16 & 13.98 & 9.55 & 0.95 & \textless{}0.01\\ \specialrule{0.5pt}{1pt}{1pt}
\multirow{3}{*}{\#18} & G-Fuzz     & \textbf{8} & \textbf{17.45} & - & - & -\\ \cline{2-7}
 & Syz-Go & 7 & 20.42 & 1.17 & 0.56 & 0.252\\ \cline{2-7}
 & Syzkaller & 6 & 19.32 & 1.11 & 0.56 & 0.247\\ \specialrule{0.5pt}{1pt}{1pt}
\multirow{3}{*}{\#19} & G-Fuzz     & \textbf{16} & \textbf{8.35} & - & - & -\\ \cline{2-7}
 & Syz-Go & 13 & 14.23 & 1.70 & 0.74 & \textless{}0.01\\ \cline{2-7}
 & Syzkaller & 12 & 16.29 & 1.95 & 0.73 & \textless{}0.01\\ \specialrule{0.5pt}{1pt}{1pt}
\multirow{3}{*}{\#20} & G-Fuzz     & \textbf{19} & \textbf{2.89} & - & - & -\\ \cline{2-7}
 & Syz-Go & 8 & 18.78 & 6.51 & 0.93 & \textless{}0.01\\ \cline{2-7}
 & Syzkaller & 16 & 14.39 & 4.99 & 0.92 & \textless{}0.01\\ \specialrule{0.5pt}{1pt}{1pt}
\multirow{3}{*}{\#21} & G-Fuzz     & \textbf{14} & \textbf{10.20} & - & - & -\\ \cline{2-7}
 & Syz-Go & 4 & 20.68 & 2.03 & 0.78 & \textless{}0.01\\ \cline{2-7}
 & Syzkaller & 12 & 17.07 & 1.67 & 0.73 & \textless{}0.01\\ \specialrule{0.5pt}{1pt}{1pt}
\multirow{3}{*}{\#22} & G-Fuzz     & \textbf{15} & \textbf{12.16} & - & - & -\\ \cline{2-7}
 & Syz-Go & 7 & 20.39 & 1.68 & 0.76 & \textless{}0.01\\ \cline{2-7}
 & Syzkaller & 9 & 18.16 & 1.49 & 0.73 & \textless{}0.01\\ \specialrule{0.5pt}{1pt}{1pt}
\multirow{3}{*}{\#23} & G-Fuzz     & \textbf{17} & \textbf{14.92} & - & - & -\\ \cline{2-7}
 & Syz-Go & 8 & 19.64 & 1.32 & 0.72 & \textless{}0.01\\ \cline{2-7}
 & Syzkaller & 11 & 17.94 & 1.20 & 0.64 & 0.069\\ \specialrule{0.5pt}{1pt}{1pt}
\multirow{3}{*}{\#24} & G-Fuzz     & \textbf{20} & \textbf{0.31} & - & - & -\\ \cline{2-7}
 & Syz-Go & 10 & 17.68 & 57.17 & 1.00 & \textless{}0.01\\ \cline{2-7}
 & Syzkaller & 15 & 13.54 & 43.81 & 1.00 & \textless{}0.01\\ \specialrule{0.5pt}{1pt}{1pt}

\multirow{3}{*}{{\#25}} & G-Fuzz     & \textbf{3} & \textbf{21.46} & - & - & -\\ \cline{2-7}
 & Syz-Go & 1 & 23.90 & 1.11 & 0.55 & 0.138\\ \cline{2-7}
 & Syzkaller & 0 & 24.00 & 1.12 & 0.57 & 0.040\\ \specialrule{0.5pt}{1pt}{1pt}
 
\multirow{3}{*}{{\#26}} & G-Fuzz     & \textbf{2} & \textbf{23.52} & - & - & -\\ \cline{2-7}
 & Syz-Go & 1 & 23.94 & 1.02 & 0.53 & 0.267\\ \cline{2-7}
 & Syzkaller & 0 & 24.00 & 1.02 & 0.55 & 0.081\\ \specialrule{0.5pt}{1pt}{1pt}
\end{tabular}}}
\end{table}

\begin{table}[!htbp]
	\centering 
	\footnotesize
	\caption{The performance of \sys, Syz-Go and Syzkaller in patch testing.}\label{tab:patch-test}
 	\renewcommand{\arraystretch}{1.12}
\setlength{\tabcolsep}{6pt}{
\scalebox{0.9}{
			\begin{tabular}{clccccc}
				\hline
\textbf{Patch ID} & \textbf{Fuzzer} & \textbf{runs} & \textbf{$\mu$TTE (h)} & \textbf{Speedup}    & \textbf{$\hat{A}_{12}$}  & \textbf{p-value} \\ \hline

\multirow{3}{*}{\#1} & G-Fuzz     & 20 & \textbf{1.16} & - & - & -\\ \cline{2-7}
 & Syz-Go & 20 & 3.41 & 2.95 & 0.83 & \textless{}0.01\\ \cline{2-7}
 & Syzkaller & 20 & 2.11 & 1.82 & 0.74 & \textless{}0.01\\ \specialrule{0.5pt}{1pt}{1pt}
\multirow{3}{*}{\#2} & G-Fuzz     & 20 & \textbf{1.01} & - & - & -\\ \cline{2-7}
 & Syz-Go & 20 & 2.40 & 2.38 & 0.77 & \textless{}0.01\\ \cline{2-7}
 & Syzkaller & 20 & 2.20 & 2.18 & 0.78 & \textless{}0.01\\ \specialrule{0.5pt}{1pt}{1pt}
\multirow{3}{*}{\#3} & G-Fuzz     & \textbf{20} & \textbf{0.07} & - & - & -\\ \cline{2-7}
 & Syz-Go & \textbf{20} & 4.42 & 61.94 & 1.00 & \textless{}0.01\\ \cline{2-7}
 & Syzkaller & 18 & 6.90 & 96.73 & 1.00 & \textless{}0.01\\ \specialrule{0.5pt}{1pt}{1pt}
\multirow{3}{*}{\#4} & G-Fuzz     & \textbf{18} & \textbf{3.75} & - & - & -\\ \cline{2-7}
 & Syz-Go & 14 & 17.11 & 4.56 & 0.89 & \textless{}0.01\\ \cline{2-7}
 & Syzkaller & 3 & 21.99 & 5.86 & 0.93 & \textless{}0.01\\ \specialrule{0.5pt}{1pt}{1pt}
\multirow{3}{*}{\#5} & G-Fuzz     & \textbf{18} & \textbf{6.08} & - & - & -\\ \cline{2-7}
 & Syz-Go & 15 & 14.50 & 2.38 & 0.85 & \textless{}0.01\\ \cline{2-7}
 & Syzkaller & 12 & 14.98 & 2.46 & 0.77 & \textless{}0.01\\ \specialrule{0.5pt}{1pt}{1pt}
\multirow{3}{*}{\#6} & G-Fuzz     & \textbf{18} & \textbf{12.06} & - & - & -\\ \cline{2-7}
 & Syz-Go & 7 & 19.98 & 1.66 & 0.81 & \textless{}0.01\\ \cline{2-7}
 & Syzkaller & 2 & 22.98 & 1.91 & 0.91 & \textless{}0.01\\ \specialrule{0.5pt}{1pt}{1pt}
\multirow{3}{*}{\#7} & G-Fuzz     & \textbf{20} & \textbf{1.24} & - & - & -\\ \cline{2-7}
 & Syz-Go & \textbf{20} & 5.75 & 4.64 & 0.91 & \textless{}0.01\\ \cline{2-7}
 & Syzkaller & 17 & 12.03 & 9.72 & 0.98 & \textless{}0.01\\ \specialrule{0.5pt}{1pt}{1pt}
\multirow{3}{*}{\#8} & G-Fuzz     & 17 & \textbf{5.24} & - & - & -\\ \cline{2-7} 
 & Syz-Go & \textbf{19} & 6.20 & 1.18 & 0.73 & \textless{}0.01\\ \cline{2-7}
 & Syzkaller & 15 & 14.05 & 2.68 & 0.85 & \textless{}0.01\\ \specialrule{0.5pt}{1pt}{1pt}
\multirow{3}{*}{\#9} & G-Fuzz     & 20 & \textbf{0.33} & - & - & -\\ \cline{2-7}
 & Syz-Go & 20 & 2.75 & 8.42 & 1.00 & \textless{}0.01\\ \cline{2-7}
 & Syzkaller & 20 & 4.28 & 13.09 & 1.00 & \textless{}0.01\\ \specialrule{0.5pt}{1pt}{1pt}
\multirow{3}{*}{\#10} & G-Fuzz     & 20 & \textbf{0.35} & - & - & -\\ \cline{2-7}
 & Syz-Go & 20 & 1.54 & 4.45 & 0.95 & \textless{}0.01\\ \cline{2-7}
 & Syzkaller & 20 & 1.80 & 5.19 & 0.94 & \textless{}0.01\\ \specialrule{0.5pt}{1pt}{1pt}
\multirow{3}{*}{\#11} & G-Fuzz     & 19 & 2.81 & - & - & -\\ \cline{2-7}
 & Syz-Go & \textbf{20} & 1.52 & 0.54 & 0.49 & 0.473\\ \cline{2-7}
 & Syzkaller & \textbf{20} & \textbf{1.50} & 0.53 & 0.49 & 0.473\\ \specialrule{0.5pt}{1pt}{1pt}
\multirow{3}{*}{\#12} & G-Fuzz     & 18 & 4.53 & - & - & -\\ \cline{2-7}
 & Syz-Go & \textbf{20} & \textbf{2.60} & 0.57 & 0.62 & 0.104\\ \cline{2-7}
 & Syzkaller & \textbf{20} & 4.67 & 1.03 & 0.69 & 0.023\\ \specialrule{0.5pt}{1pt}{1pt}
\multirow{3}{*}{\#13} & G-Fuzz     & 20 & \textbf{0.19} & - & - & -\\ \cline{2-7}
 & Syz-Go & 20 & 2.25 & 11.86 & 1.00 & \textless{}0.01\\ \cline{2-7}
 & Syzkaller & 20 & 3.68 & 19.40 & 1.00 & \textless{}0.01\\ \specialrule{0.5pt}{1pt}{1pt}
\multirow{3}{*}{\#14} & G-Fuzz     & 20 & \textbf{0.42} & - & - & -\\ \cline{2-7}
 & Syz-Go & 20 & 1.23 & 2.92 & 0.83 & \textless{}0.01\\ \cline{2-7}
 & Syzkaller & 20 & 1.52 & 3.60 & 0.92 & \textless{}0.01\\ \specialrule{0.5pt}{1pt}{1pt}
\multirow{3}{*}{\#15} & G-Fuzz     & 20 & \textbf{0.13} & - & - & -\\ \cline{2-7}
 & Syz-Go & 20 & 1.98 & 15.39 & 1.00 & \textless{}0.01\\ \cline{2-7}
 & Syzkaller & 20 & 5.25 & 40.73 & 1.00 & \textless{}0.01\\ \specialrule{0.5pt}{1pt}{1pt}
\multirow{3}{*}{\#16} & G-Fuzz     & \textbf{20} & \textbf{0.26} & - & - & -\\ \cline{2-7}
 & Syz-Go & \textbf{20} & 2.91 & 11.08 & 0.99 & \textless{}0.01\\ \cline{2-7}
 & Syzkaller & 19 & 6.32 & 24.05 & 1.00 & \textless{}0.01\\ \specialrule{0.5pt}{1pt}{1pt}
\multirow{3}{*}{\#17} & G-Fuzz     & \textbf{20} & \textbf{4.60} & - & - & -\\ \cline{2-7}
 & Syz-Go & 13 & 13.73 & 2.98 & 0.81 & \textless{}0.01\\ \cline{2-7}
 & Syzkaller & 11 & 17.33 & 3.76 & 0.89 & \textless{}0.01\\ \specialrule{0.5pt}{1pt}{1pt}
\multirow{3}{*}{\#18} & G-Fuzz     & 20 & \textbf{0.21} & - & - & -\\ \cline{2-7}
 & Syz-Go & 20 & 4.62 & 21.85 & 1.00 & \textless{}0.01\\ \cline{2-7}
 & Syzkaller & 20 & 3.53 & 16.70 & 1.00 & \textless{}0.01\\ \specialrule{0.5pt}{1pt}{1pt}
\multirow{3}{*}{\#19} & G-Fuzz     & 20 & \textbf{1.47} & - & - & -\\ \cline{2-7}
 & Syz-Go & 20 & 3.98 & 2.70 & 0.76 & \textless{}0.01\\ \cline{2-7}
 & Syzkaller & 20 & 3.90 & 2.65 & 0.82 & \textless{}0.01\\ \specialrule{0.5pt}{1pt}{1pt}
\multirow{3}{*}{\#20} & G-Fuzz     & 20 & \textbf{1.26} & - & - & -\\ \cline{2-7}
 & Syz-Go & 20 & 5.57 & 4.42 & 0.94 & \textless{}0.01\\ \cline{2-7}
 & Syzkaller & 20 & 5.92 & 4.70 & 0.94 & \textless{}0.01\\ \specialrule{0.5pt}{1pt}{1pt}

\multirow{3}{*}{\#21} & G-Fuzz     & 5 & 21.07 & - & - & -\\ \cline{2-7}
 & Syz-Go & \textbf{7} & \textbf{21.01} & 1.00 & 0.46 & 0.297\\ \cline{2-7}
 & Syzkaller & 0 & 24.00 & 1.14 & 0.62 & \textless{}0.01\\ \specialrule{0.5pt}{1pt}{1pt}

\multirow{3}{*}{\#22} & G-Fuzz     & \textbf{3} & \textbf{22.85} & - & - & -\\ \cline{2-7}
 & Syz-Go & 0 & 24.00 & 1.05 & 0.57 & 0.040\\ \cline{2-7}
 & Syzkaller & 0 & 24.00 & 1.05 & 0.57 & 0.040\\ \specialrule{0.5pt}{1pt}{1pt}

	\end{tabular}}}
\end{table}

\FloatBarrier 
\subsection{Experimental Settings}

 \textbf{Compared Fuzzers.}
We compare \sys with two fuzzers: Syzkaller and Syz-Go.
Syzkaller is the state-of-the-art kernel fuzzer.
We implement Syz-Go by adopting the method of \aflgo on Syzkaller.
Specifically, as the time overhead of the original \aflgo implementation in distance calculation is high, we use the same distance calculation implementation in Syz-Go as \sys.

For the directed fuzzing part of Syz-Go, we implement the simulated annealing-based power scheduling algorithm of \aflgo in it.
Following the original design of \aflgo~\cite{ma2011directed}, for the 24-hour experiment, the exploration time of Syz-Go is 20 hours, and the exploitation time is 4 hours.
For the 72-hour experiment, the exploration time of Syz-Go is 60 hours, and the exploitation is 12 hours.

\textbf{Evaluation Metrics.}
Following the methods of the state-of-the-art directed fuzzing papers~\cite{bohme2017directed, chen2018hawkeye}, we mainly use \emph{runs} and \emph{Time-to-Exposure (TTE)} as the evaluation metrics.
Specifically, \emph{runs} represents the times that a fuzzer triggers the target in multiple repeated experiments.
\emph{TTE} is the first time that a fuzzer triggers the target, and \emph{$\mu$TTE} is the arithmetic average of \emph{TTE} in multiple repeated experiments.
In specific, if a fuzzer does not trigger the target over the fuzzing timeout (e.g., 24 hours), the \emph{TTE} of this experiment is regarded as the timeout.

Moreover, we use statistical metrics such as $\hat{A}_{12}$~\cite{vargha2000critique} and p-value to provide more comprehensive evaluations.
$\hat{A}_{12}$ is the metric to measure the effect size, that is, the probability of one group of samples perform better than the other.
The p-value is to measure whether the difference between the two groups is real or due to randomness.
If the p-value is less than 0.05 (or 0.01), there is a significant difference between the two groups.
We use the Mann-Whitney U test, a non-parametric statistical test, to calculate the p-value.

 \textbf{Initial Seeds and Environment.}
For Syzkaller and Syz-Go, the initial seeds are the same, generated by the original design of Syzkaller.
For \sys, the initial seeds are generated according to the inferred target related syscalls.
We conduct all the experiments on several servers with the same settings.
Each server has 20 Intel Xeon E5-2650 V4 CPUs, and the OS is 64-bits Ubuntu 16.04 LTS.

\subsection{General Target Testing}

In this scenario, the targets are the arbitrary locations of \gv. 
We randomly select 50 code sites as the targets.
Among the 50 targets, there are 24 targets that no fuzzer can find with the 24-hour $\times$ 20 repetition evaluation.
Table~\ref{tab:normal-target} shows the performance of \gv, Syz-Go and Syzkaller in directed fuzzing for the 26 targets.

From Table~\ref{tab:normal-target}, we can get the following observations.
(1) It is evident that \sys outperform both Syzkaller and Syz-Go in discovering the targets.
In terms of $\hat{A}_{12}$ metric,
\sys achieves more than 0.5 $\hat{A}_{12}$ values on all targets comparing to Syz-Go and Syzkaller.
Specifically, \sys achieves more than 0.71 $\hat{A}_{12}$ on 21 targets, except for target \#7, \#11, \#18, \#25 and \#26.
$\hat{A}_{12}$ value is more than 0.71 means that the probability of one group performs better than the other is large~\cite{vargha2000critique}.
For 20 targets, the p-value of \sys between Syz-Go and Syzkaller is less than 0.01, demonstrating the significantly better performance of \sys.
(2) \sys has more stable performance than Syz-Go and Syzkaller.
According to the \emph{runs} column, \sys has the highest values in 25 targets, except for target \#4.
(3) \sys accelerates the performance of directed fuzzing by a large margin.
For instance, \sys only takes 0.92 hours in triggering target \#2, whereas Syz-Go and Syzkaller need to use more than 12 hours.
More importantly, for target \#25 and \#26, G-Fuzz can trigger them while Syzkaller cannot.
From the \emph{speedup} column, we can see that the max \emph{speedup} that \sys can achieve is more than 130 (target \#3).
For seven targets, the \emph{speedup} value of \sys is more than 10.
In summary, the performance of \sys in testing the general targets of \gv is stable and efficient.

\subsection{Patch Testing}

The targets of patch testing are the code modified by the patch commits.
In this scenario, we select 50 code locations modified by the patch commits as the targets.
Among the 50 patch targets, 28 of them are not found by any fuzzer with the 24-hour $\times$ 20 repetition evaluation.
Table~\ref{tab:patch-test} shows the performance of \sys, Syzkaller and Syz-Go in directed fuzzing for the rest of 22 patch targets.

From Table~\ref{tab:patch-test}, we can get the following observations.
(1) Under the \emph{runs} metrics, \sys has the best performance among the compared fuzzers on 18 targets, which demonstrates the effectiveness of \sys.
(2) In terms of $\mu$TTE and \emph{speedup} metrics, \sys takes the least amount of time in triggering 20 targets, except for target \#11 and \#12.
Moreover, \sys can boost the performance of patch testing up to 96.73 times.
(3) As for the $\hat{A}_{12}$ metric, \sys has more than 0.5 $\hat{A}_{12}$ value on 20 targets and has more than 0.71 $\hat{A}_{12}$ value on 18 targets.
(4) For over 90 percent of targets (18/20), the p-value of \sys is less than 0.01. 
In summary, \sys outperforms Syz-Go and Syzkaller in patch testing.

\begin{table}[!h]
    \centering 
    \footnotesize
    \caption{The performance of \sys, Syz-Go and Syzkaller in bug reproduction.}\label{tab:bug-test}
    \setlength{\tabcolsep}{6pt}{
        \scalebox{0.9}{
        \begin{tabular}{clccccc}
\hline
         \textbf{Bug ID} & \textbf{Fuzzer} & \textbf{runs} & \textbf{$\mu$TTE (h)} & \textbf{Speedup} & \textbf{$\hat{A}_{12}$} & \textbf{p-value} \\ \hline

\multirow{3}{*}{\#1} & G-Fuzz     & \textbf{20} & \textbf{3.91} & - & - & -\\ \cline{2-7}
 & Syz-Go & \textbf{20} & 13.53 & 3.46 & 0.88 & \textless{}0.01\\ \cline{2-7}
 & Syzkaller & 19 & 15.49 & 3.96 & 0.85 & \textless{}0.01\\ \specialrule{0.5pt}{1pt}{1pt}
\multirow{3}{*}{\#2} & G-Fuzz     & 20 & \textbf{0.08} & - & - & -\\ \cline{2-7}
 & Syz-Go & 20 & 0.84 & 10.74 & 0.99 & \textless{}0.01\\ \cline{2-7}
 & Syzkaller & 20 & 2.09 & 26.82 & 1.00 & \textless{}0.01\\ \specialrule{0.5pt}{1pt}{1pt}
\multirow{3}{*}{\#3} & G-Fuzz     & \textbf{3} & \textbf{65.37} & - & - & -\\ \cline{2-7}
 & Syz-Go & 2 & 66.68 & 1.02 & 0.52 & 0.345\\ \cline{2-7}
 & Syzkaller & \textbf{3} & 69.71 & 1.07 & 0.51 & 0.431\\ \specialrule{0.5pt}{1pt}{1pt}
\multirow{3}{*}{\#4} & G-Fuzz     & \textbf{8} & \textbf{61.87} & - & - & -\\ \cline{2-7}
 & Syz-Go & 1 & 70.32 & 1.14 & 0.67 & \textless{}0.01\\ \cline{2-7}
 & Syzkaller & 1 & 71.22 & 1.15 & 0.68 & \textless{}0.01\\ \specialrule{0.5pt}{1pt}{1pt}
\multirow{3}{*}{\#5} & G-Fuzz     & \textbf{4} & \textbf{62.16} & - & - & -\\ \cline{2-7}
 & Syz-Go & 2 & 69.61 & 1.12 & 0.56 & 0.175\\ \cline{2-7}
 & Syzkaller & 1 & 71.70 & 1.15 & 0.58 & 0.069\\ \specialrule{0.5pt}{1pt}{1pt}
\multirow{3}{*}{\#6} & G-Fuzz     & \textbf{3} & \textbf{67.50} & - & - & -\\ \cline{2-7}
 & Syz-Go & 1 & 69.72 & 1.03 & 0.55 & 0.162\\ \cline{2-7}
 & Syzkaller & 0 & 72.00 & 1.07 & 0.57 & 0.040\\ \specialrule{0.5pt}{1pt}{1pt}
\multirow{3}{*}{\#7} & G-Fuzz     & \textbf{6} & 63.30 & - & - & -\\ \cline{2-7}
 & Syz-Go & 5 & \textbf{60.60} & 0.96 & 0.51 & 0.466\\ \cline{2-7}
 & Syzkaller & 3 & 65.75 & 1.04 & 0.57 & 0.163\\ \specialrule{0.5pt}{1pt}{1pt}
\multirow{3}{*}{\#8} & G-Fuzz     & \textbf{17} & \textbf{15.42} & - & - & -\\ \cline{2-7}
 & Syz-Go & 10 & 48.87 & 3.17 & 0.84 & \textless{}0.01\\ \cline{2-7}
 & Syzkaller & 5 & 62.27 & 4.04 & 0.89 & \textless{}0.01\\ \specialrule{0.5pt}{1pt}{1pt}
\multirow{3}{*}{\#9} & G-Fuzz     & \textbf{7} & \textbf{52.44} & - & - & -\\ \cline{2-7}
 & Syz-Go & 0 & 72.00 & 1.37 & 0.68 & \textless{}0.01\\ \cline{2-7}
 & Syzkaller & 1 & 70.48 & 1.34 & 0.66 & \textless{}0.01\\ \specialrule{0.5pt}{1pt}{1pt}
\multirow{3}{*}{\#10} & G-Fuzz     & \textbf{6} & \textbf{57.73} & - & - & -\\ \cline{2-7}
 & Syz-Go & 0 & 72.00 & 1.25 & 0.65 & \textless{}0.01\\ \cline{2-7}
 & Syzkaller & 2 & 68.24 & 1.18 & 0.60 & 0.054\\ \specialrule{0.5pt}{1pt}{1pt}

\multirow{3}{*}{\#11} & G-Fuzz     & \textbf{8} & \textbf{56.93} & - & - & -\\ \cline{2-7}
 & Syz-Go & 5 & 63.46 & 1.11 & 0.57 & 0.186\\ \cline{2-7}
 & Syzkaller & 0 & 72.00 & 1.26 & 0.70 & \textless{}0.01\\ \specialrule{0.5pt}{1pt}{1pt}
 
\end{tabular}}}
\end{table}

\begin{table*}[htbp]
\centering 
 \footnotesize
\caption{The time overhead of \aflgo and G-Fuzz in distance calculation.}\label{tab:gfuzz-dis}
\setlength{\tabcolsep}{9pt}{
\scalebox{0.9}{
\begin{tabular}{l|l|c|c|c}
\hline
\multirow{2}{*}{\textbf{Commit   ID}} & \multicolumn{1}{c|}{\multirow{2}{*}{\textbf{Target}}} & \multirow{2}{*}{\textbf{Reachable Function Ratio}} & \multicolumn{2}{c}{\textbf{Time(s) of Static Distance Calculation}} \\ \cline{4-5} 
&  &  & \textbf{AFLGo}      & \textbf{G-Fuzz} \\ \hline
cdf49c44 & pkg/tcpip/stack/linkaddrcache.go:189 & 18.88\% & 76,275 & 102.0 \\ \hline
c564293b6 & pkg/sentry/syscalls/linux/vfs2/splice.go:398 & 0.16\% & 60,507 & 92.8 \\ \hline
76da673a & pkg/tcpip/network/ipv4/igmp.go:158 & 18.90\% & 76,233 & 102.6 \\ \hline
8b9cb36d1 & pkg/sentry/syscalls/linux/vfs2/splice.go:147 & 0.16\% & 60,339 & 91.6 \\ \hline
55332aca9 & pkg/sentry/vfs/file\_description.go:828 & 0.17\% & 75,937 & 103.0 \\ \hline
7f89a26e1 & pkg/sentry/kernel/fd\_table.go:369 & 0.19\% & 60,782 & 90.8 \\ \hline
\multicolumn{3}{c|}{\textbf{Average}} & 68,346 & 97.1 \\ \hline
\end{tabular}}}
\end{table*}

\subsection{Bug Reproduction}
Bug reproduction is a significant application scenario of directed fuzzing.
Assuming that we have already known the information about the bug, such as the vulnerable locations or the stack trace, etc., but do not have the PoC to validate the bug.
The goal of directed fuzzing is to generate the PoC that can trigger the bug.
In this scenario, we select the public \gv bugs as the targets from the official Syzbot website~\cite{syzbot-gVisor}.
For the public bugs, we first use the provided PoCs to validate whether the bugs can be reproduced.
Finally, we select 30 bugs as the targets.
It needs to be emphasized that we only use the provided PoCs when selecting target bugs.
In the experiment, we assume that we do not have the PoCs and the goal of directed fuzzing is to generate the PoCs.
For the 30 bugs, only 11 of them can be successfully triggered by at least one fuzzer within 72 hours with 20 repetitions.
Table~\ref{tab:bug-test} presents the performance of \sys, Syz-Go and Syzkaller in bug reproduction.

From Table~\ref{tab:bug-test}, we can observe the following facts.
(1) For 11 bugs, \sys successfully trigger all of them, whereas Syz-Go fails to trigger two of them, and Syzkaller fails to trigger two of them.
(2) In terms of the \emph{runs} metric, \sys achieves the highest number for all bugs, which is significantly better than Syz-Go and Syzkaller.
(3) The speed of \sys in discovering the bugs is faster than both Syz-Go and Syzkaller.
According to the \emph{speedup} metric, \sys can boost the performance of bug reproduction up to 26.82 (target \#2).
(4) Compared to Syz-Go and Syzkaller, \sys has over 0.5 $\hat{A}_{12}$ value in reproducing all 11 bugs.
Besides, for seven bugs, the p-value of \sys is less than 0.01.
In summary, the above observations prove that \sys has outstanding performance in reproducing bugs.

\section{Further Analysis}\label{sec:further}

To provide comprehensive evaluations, in this section, we conduct further experiments and analysis on the effectiveness of each core methods of \sys, as well as ablation studies.
Moreover, we present the deployment and the application of \sys in real-world.

\subsection{Static Distance Calculation Method}\label{subsec:distance}

Below, we measure the performance of the static distance calculation method of \sys from two aspects: time overhead and precision.

\textbf{Time Overhead.}
To measure the time overhead of \sys in distance calculation, we randomly select six versions of \gv. 
We randomly select a code location as the target for each version to calculate the static distance for the rest code of \gv. 
We compare the \sys with \aflgo. 
Table~\ref{tab:gfuzz-dis} presents the average time overhead of \aflgo and \sys in static distance calculation with three repetitions.
For the six targets, on average, \aflgo spends 68,346 seconds (18.9 hours) in calculating the static distance, while \sys only spends 97.1 seconds.
Compared to \aflgo, \sys reduces the time overhead in distance calculation drastically.
The third column of Table~\ref{tab:gfuzz-dis} shows the reachable function ratio of each target.
Among the six targets, the highest ratio is 18.90\%, and four of them only have less than 0.2\% reachable ratios.
Note that as we use \emph{RTA} method to identify the indirect calls, there is almost no false negative in the result and the number of ground truth value of reachable function ratio may be even smaller/less. 
This observation shows that the target related code only accounts for a tiny fraction.
The time of calculating static distance for the unrelated code is unnecessary and wasteful.
Thus, \sys's reachability analysis based distance calculation can effectively reduce the time overhead.

In addition, we conduct a time profiling experiment on \aflgo.
We find that the \texttt{find\_nodes} function of \aflgo takes more than 50\% time, and this function has been invoked almost two billion times on average.
Given a string of a basic block (i.e., file name and the line number), the functionality of \texttt{find\_nodes} function is to find its corresponding node object in the current CFG.
Nevertheless, \texttt{find\_nodes} function will be called for each basic block of all functions, whereas most of the traversed basic blocks do not belong to the current CFG.
The complexity of this implementation is $O(N^{2})$, where N is the number of basic blocks.
The complexity can be reduced to $O(N)$ by only traversing basic blocks of the current CFG.

\textbf{Precision of Distance.}
In addition to reducing the time overhead in distance calculation, \sys improves the distance's precision, compared with \aflgo.
Based on the discussion in \S \ref{subsec:gfuzz-distance}, there are two main issues of \aflgo in the precision of distance: (1) \aflgo does not consider the indirect calls.
(2) \aflgo uses the coarse-grained function level distance to approximate the basic block level distance.
Next, we discuss the impact of these two issues on directed fuzzing and whether \sys can solve them.

\begin{figure}[!t]
    \centering
    \hspace*{-3mm}\includegraphics[width=3.6in]{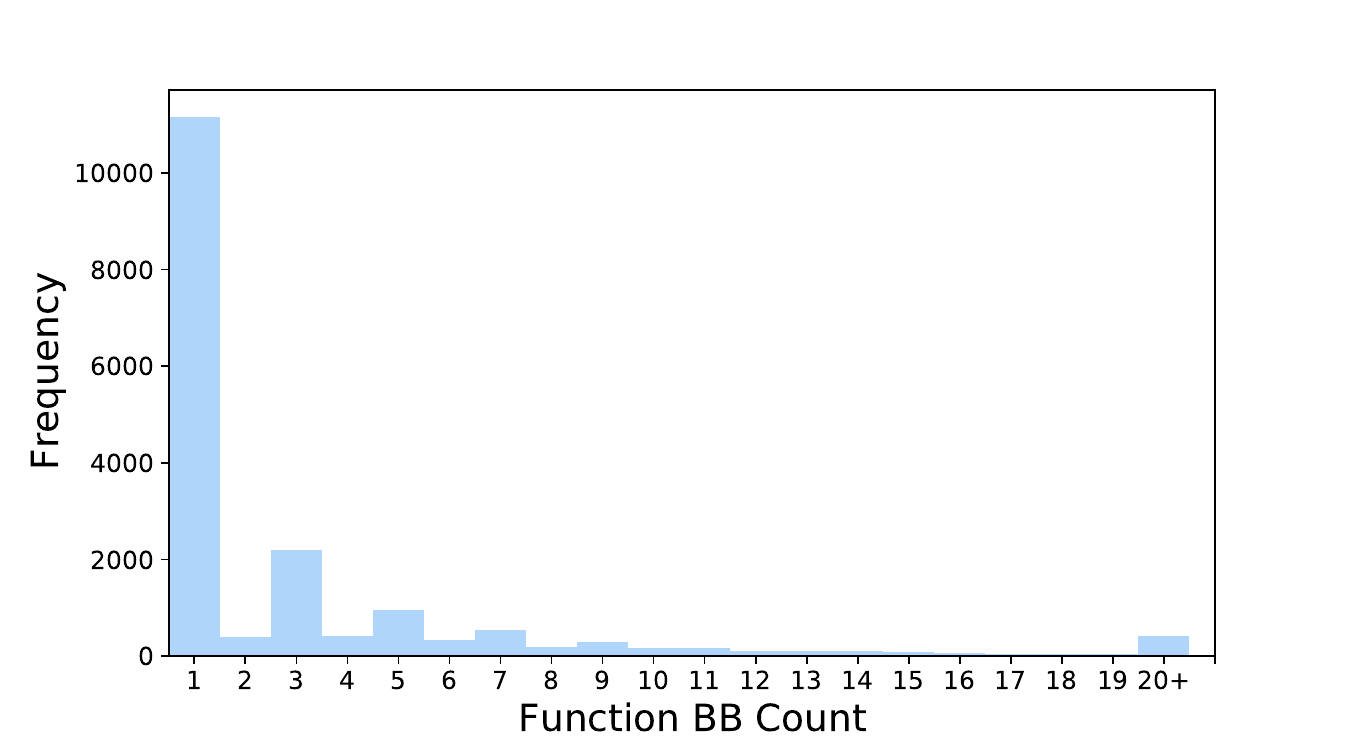}
    \caption{The frequency distribution of the number of basic blocks within a function of \gv.}
    \label{fig:bbhist}
     \vspace{-5mm}
\end{figure}

For issue (1), Table~\ref{tab:gfuzz-indirect} shows the number of indirect calls of \gv with six different versions.
We can observe that the ratio of the indirect calls to the total calls ranges from 4.89\% to 5.24\%.
Although the ratio of the indirect calls is not very large, it may cause missing edges.
In particular, the absence of some critical edges will cause many errors in distance calculation.
The sixth and seventh columns of Table~\ref{tab:gfuzz-indirect} lists the number of target related functions (i.e., have the static distance) that can be found by \aflgo and \sys.
We can observe that the number of target related functions decreases by a large margin if the indirect calls are not considered, which makes the fuzzer fail to find some important seeds.
For issue (2), we present frequency distribution of the number of basic blocks within a \gv function in Fig.~\ref{fig:bbhist}. 
We can observe that the number of basic blocks has a wide range.
Most of the \gv functions have only one basic block.
Some functions even have more than 20 basic blocks.
The largest number of basic blocks that a \gv function has is 162.
Thus, instead of using a constant to approximate, we provide a fine-grained (i.e., basic block level) distance information in the implementation of \sys.
Moreover, compared to \aflgo,
\sys uses \emph{RTA} to discover the indirect calls and calculates the basic block level distance with a low time overhead, effectively solves the two issues.

\begin{table*}[tbp]
    \centering 
    \footnotesize
    \caption{The impact of the indirect calls on the reachable sets of \gv.}\label{tab:gfuzz-indirect}
    \setlength{\tabcolsep}{6pt}{
        \scalebox{0.9}{
    \begin{tabular}{l|l|c|c|c|c|c}
\hline
\multicolumn{1}{c|}{\multirow{2}{*}{\textbf{Commit   ID}}} & \multicolumn{1}{c|}{\multirow{2}{*}{\textbf{Target}}} & \multicolumn{1}{c|}{\multirow{2}{*}{\textbf{\#Call Sites}}} & \multicolumn{1}{c|}{\multirow{2}{*}{\textbf{\#Indirect Call}}} & \multicolumn{1}{c|}{\multirow{2}{*}{\textbf{Indirect Call Ratio}}} & \multicolumn{2}{c}{\textbf{\# Target Related Functions}} \\ \cline{6-7}
\multicolumn{1}{c|}{} & \multicolumn{1}{c|}{} & \multicolumn{1}{c|}{} & \multicolumn{1}{c|}{} & \multicolumn{1}{c|}{} & \textbf{\aflgo} & \textbf{G-Fuzz} \\ \hline
cdf49c44 & pkg/tcpip/stack/linkaddrcache.go:189 & 79,048 & 4,142 & 5.24\% & 12 & 3,376 \\ \hline
c564293b6 & pkg/sentry/syscalls/linux/vfs2/splice.go:398 & 70,725 & 3,467 & 4.90\% & 1 & 24 \\ \hline
76da673a & pkg/tcpip/network/ipv4/igmp.go:158 & 79,047 & 4,131 & 5.23\% & 1 & 3,376 \\ \hline
8b9cb36d1 & pkg/sentry/syscalls/linux/vfs2/splice.go:147 & 70,492 & 3,448 & 4.89\% & 1 & 24 \\ \hline
55332aca9 & pkg/sentry/vfs/file\_description.go:828 & 78,922 & 4,120 & 5.22\% & 1 & 30 \\ \hline
7f89a26e1 & pkg/sentry/kernel/fd\_table.go:369 & 70,872 & 3,476 & 4.90\% & 2 & 29 \\ \hline
\end{tabular}}}
\end{table*}

\begin{figure}[htbp]
\centering
  \includegraphics[width=0.9\linewidth]{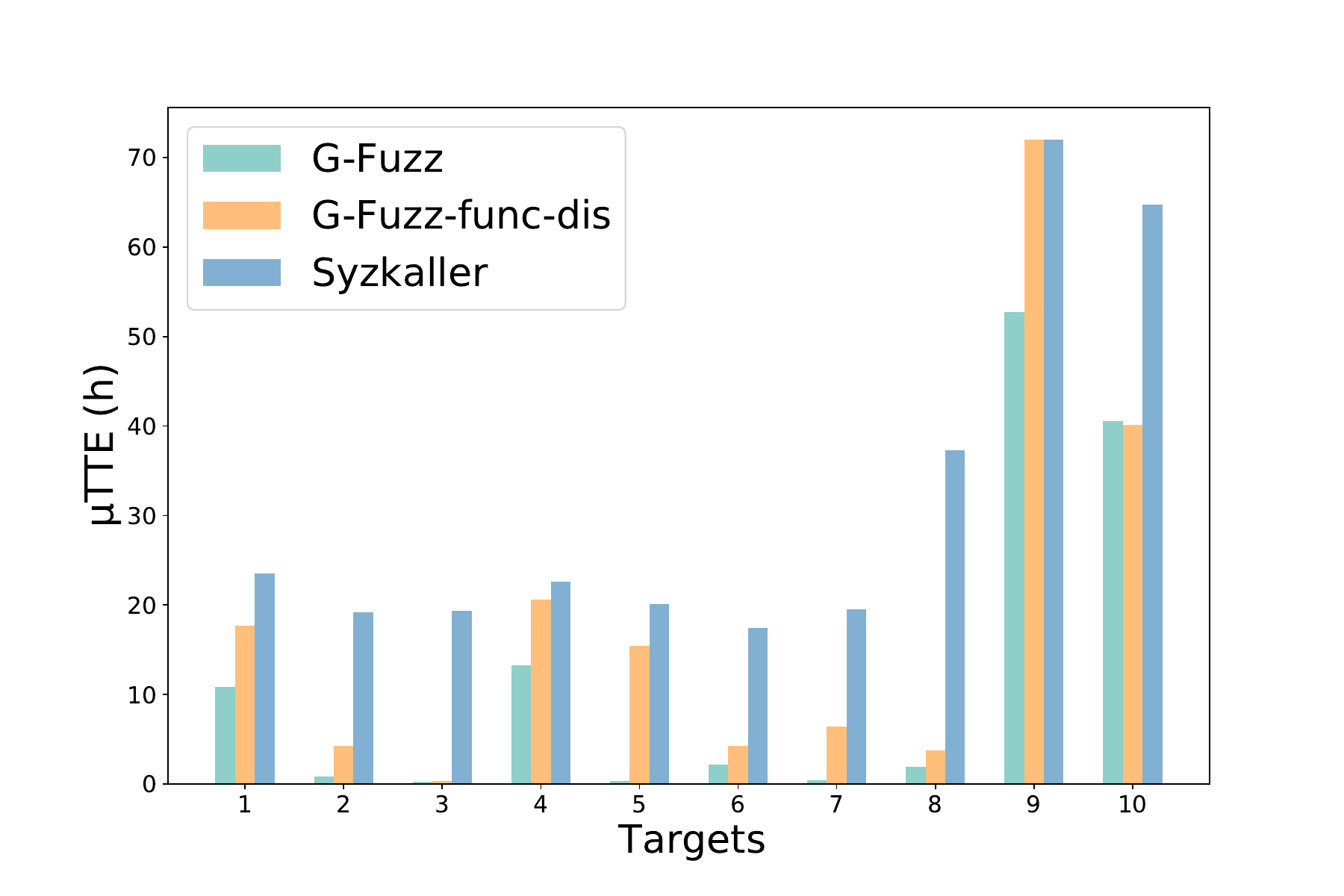}
  \caption{Ablation study 1: static distance calculation optimization.}
  \label{fig:funcdis}
  \vspace{-5mm}
\end{figure}

\textbf{Ablation Study.}
We conduct ablation study on the static distance calculation optimization.
we compare G-Fuzz to G-Fuzz-func-dis (G-Fuzz that uses function level distance multiplied by a constant 10 to approximate basic block level distance) and Syzkaller in directed fuzzing 10 targets, with five repetitions.
Fig.~\ref{fig:funcdis} shows the experimental results.
We can obverse that, as a whole, G-Fuzz uses significantly less time in triggering the targets than G-Fuzz-func-dis and Syzkaller.
G-Fuzz outperforms G-Fuzz-func-dis in 8 out of 10 targets (except for target 3 and target 10), and outperforms Syzkaller in 10 targets.
Therefore, the experimental results demonstrate the effectiveness of the distance calculation optimization of G-Fuzz in directed fuzzing.

\subsection{Target Related Syscall Inference and Utilization}

We evaluate this method from two aspects: the precision of the inference and the effectiveness of the utilization.

\textbf{Precision of Inference.}
To verify whether the inferred target related syscalls can help to trigger the targets, 
we compared the syscalls of the PoC\footnote{For the convenience of expression, here we use PoC to refer to the inputs that can trigger the target.} to the inference results.
As there may exist syscalls in the PoC that have no contribution to triggering the target, we perform \emph{minimization} operations on the PoC to remove redundant syscalls.
In particular, we delete each syscall of a PoC in turn to obtain a pruned PoC and execute it.
If the pruned PoC can still trigger the target, then the deleted syscall will be considered redundant.
We can obtain the \emph{minimized PoC} by removing all redundant syscalls.
We define the precision of the inference as the number of inferred syscalls that are contained in the \emph{minimized PoC} divided by the number of all inferred syscalls.

Table~\ref{fig:gfuzz-infer-target} and Table~\ref{fig:gfuzz-infer-patch} present the inferred syscalls and the precision for the general targets and the patch testing targets, respectively.
For the general target \#2, \#18, and patch \#11, \#22, we cannot reproduce the PoC, and their \emph{minimized PoC} is not presented in the table.
The inferred syscalls that are contained in the \emph{minimized PoC} is presented with bold.

We can observe that for most of the targets, the precision of inference is 100\%, which means that all the inferred syscalls are necessary for triggering these targets.
None of the target's inference precision is zero, which demonstrates that at least one inferred syscall could contribute to finding the target.
It needs to emphasize that \sys provides solutions for mitigating the impacts caused by the false positives and negatives of the inference.
For false positives, \sys uses a strategy (\ref{subsec:utilization}) to reduce the selection probabilities of the false inferred syscalls.
For false negatives, \sys reduces it by combining directed fuzzing with path exploration (\ref{sub:eeswitch}).
Therefore, \sys can effectively improve the performance of directed fuzzing.

\begin{figure}[t!]
    \centering
    \includegraphics[width=3.4in]{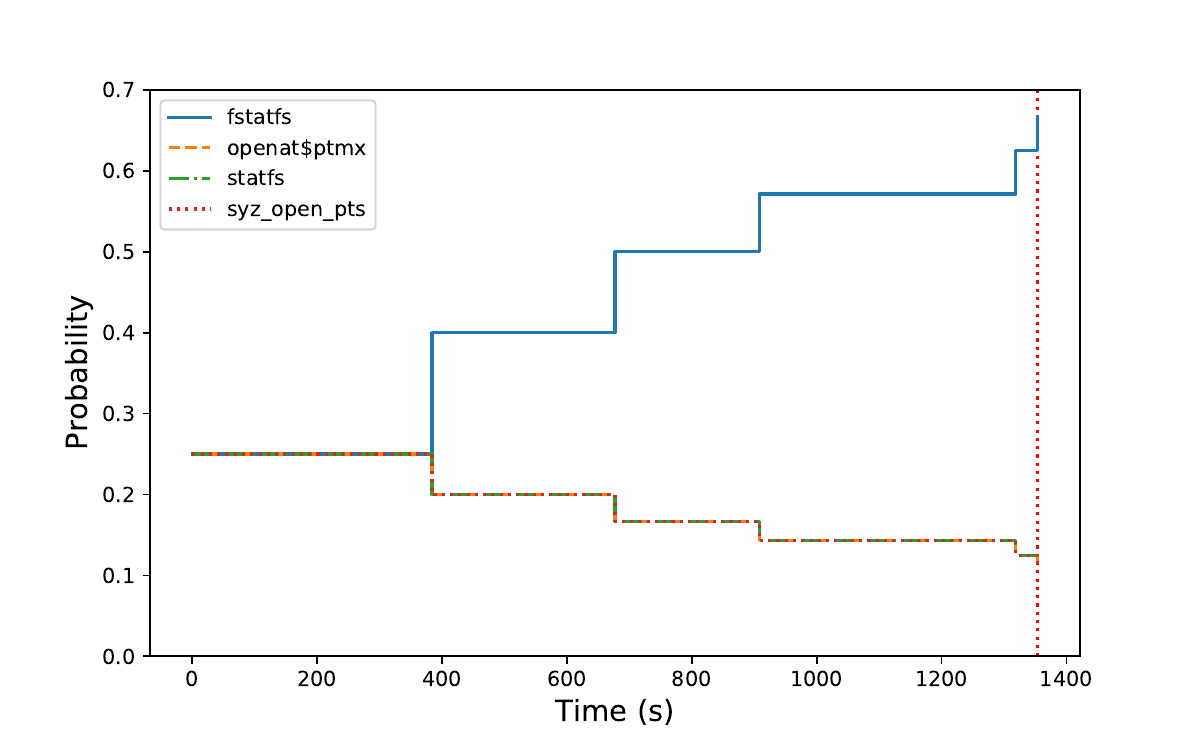}
    \caption{The change of the selection probability of each inferred syscall during the fuzzing process.}
    \label{fig:smalldis}
     \vspace{-5mm}
\end{figure}

\textbf{Effectiveness of Utilization.}
There may exist false positives in the inference results.
Thus, as we describe in \S \ref{subsec:utilization}, \sys adjusts the selection probability of each syscall dynamically according to the current fuzzing state.
To verify the effectiveness of this strategy, we record the selection probability of each inferred syscall during the fuzzing process.
Fig.~\ref{fig:smalldis} presents the curve of the selection probability of each inferred syscall of patch target \#9 with the fuzzing time increasing.
For this target, \sys infers four target related syscalls: \texttt{fstatfs}, \texttt{statfs}, \texttt{openat\$ptmx} and \texttt{syz\_open\_pts}.
From Fig.~\ref{fig:smalldis}, we can observe that the initial selection probability for the four inferred syscalls are equal.
With the fuzzing going on, \sys finds that using \texttt{fstatfs} can reduce the distance to the target effectively.
Thus, the selection probability of \texttt{fstatfs} increases, while the probabilities of the rest three syscalls decrease.
At time 1,354 seconds, \sys successfully triggers the target, which is denoted in the red line of the figure, and the PoC contains the syscall \texttt{fstatfs}.
This observation indicates that the proposed selection probability scheduling strategy can mitigate the impacts caused by the false positives of the inference.

\begin{table}[!tb]
    \centering 
    \caption{The inferred syscalls for the general targets.}
    \label{fig:gfuzz-infer-target}
    \setlength{\tabcolsep}{0pt}
    {
\renewcommand{\arraystretch}{1.2}
        \scalebox{0.85}{
 
    \begin{tabular}{>{\centering}m{0.1\textwidth}|>{\centering}m{0.35\textwidth}|>{\centering\arraybackslash}m{0.1\textwidth}}
\hline
\textbf{Target ID}   & \textbf{Inferred Syscalls}   & \textbf{Precision}     \\ \hline

\#1  & \textbf{\texttt{syz\_emit\_ethernet}}                                                                                    & 100\% \\ \hline
\#2  & \texttt{syz\_emit\_ethernet}                                                                                             & -     \\ \hline
\#3  & \textbf{\texttt{syz\_emit\_ethernet}}                                                                                    & 100\% \\ \hline
\#4  & \textbf{\texttt{syz\_emit\_ethernet}}, \textbf{\texttt{syz\_emit\_ethernet\$ipv4\_igmp}}                                 & 100\% \\ \hline
\#5  & \textbf{\texttt{syz\_emit\_ethernet}}                                                                                    & 100\% \\ \hline
\#6  & \textbf{\texttt{syz\_emit\_ethernet}}, \textbf{\texttt{syz\_emit\_ethernet\$ipv4}}                                       & 100\% \\ \hline
\#7  & \textbf{\texttt{recvmmsg}}, \textbf{\texttt{recvmsg}}                                                                    & 100\% \\ \hline
\#8  & \textbf{\texttt{tee}}                                                                                                    & 100\% \\ \hline
\#9  & \textbf{\texttt{getcwd}}                                                                                                 & 100\% \\ \hline
\#10 & \textbf{\texttt{splice}}, \textbf{\texttt{tee}}                                                                          & 100\% \\ \hline
\#11 & \textbf{\texttt{waitid}}                                                                                                 & 100\% \\ \hline
\#12 & \textbf{\texttt{link}}, \texttt{linkat}                                                                                  & 50\%  \\ \hline
\#13 & \textbf{\texttt{alarm}}                                                                                                  & 100\% \\ \hline
\#14 & \textbf{\texttt{dup}}                                                                                                    & 100\% \\ \hline
\#15 & \textbf{\texttt{mremap}}                                                                                                 & 100\% \\ \hline
\#16 & \textbf{\texttt{mbind}}, \textbf{\texttt{set\_mempolicy}}                                                                & 100\% \\ \hline
\#17 & \textbf{\texttt{unshare}}                                                                                                & 100\% \\ \hline
\#18 & \texttt{io\_destroy}                                                                                                     & -     \\ \hline
\#19 & \texttt{mmap}, \textbf{\texttt{mprotect}}, \textbf{\texttt{munmap}}, \textbf{\texttt{wait4}}                             & 75\%  \\ \hline
\#20 & \textbf{\texttt{getpeername}}, \textbf{\texttt{getsockname}}, \texttt{mmap}, \textbf{\texttt{mprotect}}, \texttt{munmap} & 60\%  \\ \hline
\#21 & \textbf{\texttt{getsockopt}}, \texttt{mmap}, \textbf{\texttt{mprotect}}, \texttt{munmap}                                 & 50\%  \\ \hline
\#22 & \textbf{\texttt{recvmmsg}}                                                                                               & 100\% \\ \hline
\#23 & \textbf{\texttt{sendmmsg}}                                                                                               & 100\% \\ \hline
\#24 & \textbf{\texttt{io\_setup}}, \texttt{mmap}, \textbf{\texttt{mprotect}}, \texttt{munmap}                                  & 50\%  \\ \hline
{\#25} & \texttt{sendfile}, \textbf{\texttt{syz\_emit\_ethernet\$ipv6}}   & 50\% \\  \hline
{\#26} & \textbf{\texttt{getsockopt}}, \texttt{syz\_emit\_ethernet} & 50\% \\  \hline
\end{tabular}}}
\end{table}

\begin{table}[!htb]
\centering 
\caption{The inferred syscalls for the patch targets.}
\label{fig:gfuzz-infer-patch}
\setlength{\tabcolsep}{0pt}{
\renewcommand{\arraystretch}{1.2}
\scalebox{0.85}{
\begin{tabular}
{>{\centering}m{0.08\textwidth}|>{\centering}m{0.35\textwidth}|>{\centering\arraybackslash}m{0.1\textwidth}}
\hline
\textbf{Patch ID}  & \textbf{Inferred Syscalls}  & \textbf{Precision}   \\ \hline
\#1  & \textbf{\texttt{getxattr}}, \textbf{\texttt{lgetxattr}}                                                                               & 100\% \\ \hline
\#2  & \textbf{\texttt{lremovexattr}}, \textbf{\texttt{removexattr}}                                                                         & 100\% \\ \hline
\#3  & \textbf{\texttt{syz\_emit\_ethernet}}, \textbf{\texttt{syz\_emit\_ethernet\$arp}}                                                     & 100\% \\ \hline
\#4  & \textbf{\texttt{syz\_emit\_ethernet}}, \textbf{\texttt{syz\_emit\_ethernet\$ipv6}}, \textbf{\texttt{syz\_emit\_ethernet\$ipv6\_icmp}} & 100\% \\ \hline
\#5  & \textbf{\texttt{socketpair}}                                                                                                          & 100\% \\ \hline
\#6  & \textbf{\texttt{syz\_emit\_ethernet}}, \textbf{\texttt{syz\_emit\_ethernet\$ipv4}}                                                    & 100\% \\ \hline
\#7  & \textbf{\texttt{syz\_emit\_ethernet}}, \textbf{\texttt{syz\_emit\_ethernet\$ipv4}}                                                    & 100\% \\ \hline
\#8  & \textbf{\texttt{syz\_emit\_ethernet}}, \textbf{\texttt{syz\_emit\_ethernet\$ipv4}}                                                    & 100\% \\ \hline
\#9  & \textbf{\texttt{fstatfs}}, \textbf{\texttt{openat\$ptmx}}, \texttt{statfs}, \textbf{\texttt{syz\_open\_pts}}                          & 75\%  \\ \hline
\#10 & \textbf{\texttt{fstatfs}}, \textbf{\texttt{pipe}}, \textbf{\texttt{pipe2}}, \texttt{statfs}                                           & 75\%  \\ \hline
\#11 & \texttt{pread64}, \texttt{preadv}, \texttt{preadv2}, \texttt{read}, \texttt{readv}                                                    & -     \\ \hline
\#12 & \textbf{\texttt{syz\_emit\_ethernet}}, \textbf{\texttt{syz\_emit\_ethernet\$ipv6}}, \textbf{\texttt{syz\_emit\_ethernet\$ipv6\_icmp}} & 100\% \\ \hline
\#13 & \textbf{\texttt{connect}}, \texttt{syz\_emit\_ethernet\$ipv4\_udp}, \texttt{syz\_emit\_ethernet\$ipv6\_udp}                           & 33\%  \\ \hline
\#14 & \textbf{\texttt{setsockopt}}                                                                                                          & 100\% \\ \hline
\#15 & \textbf{\texttt{syz\_emit\_ethernet}}, \textbf{\texttt{syz\_emit\_ethernet\$ipv4}}, \textbf{\texttt{syz\_emit\_ethernet\$ipv4\_igmp}} & 100\% \\ \hline
\#16 & \textbf{\texttt{splice}}                                                                                                              & 100\% \\ \hline
\#17 & \textbf{\texttt{splice}}                                                                                                              & 100\% \\ \hline
\#18 & \textbf{\texttt{sendfile}}                                                                                                            & 100\% \\ \hline
\#19 & \textbf{\texttt{syz\_emit\_ethernet}}                                                                                                 & 100\% \\ \hline
\#20 & \textbf{\texttt{syz\_emit\_ethernet}}, \textbf{\texttt{syz\_emit\_ethernet\$ipv6}}                                                    & 100\% \\ \hline
{\#21} & \texttt{sendfile}, \textbf{\texttt{syz\_emit\_ethernet}}   & 50\% \\  \hline
{\#22} & \texttt{sendfile}, \texttt{syz\_emit\_ethernet\$ipv4\_icmp}, \texttt{syz\_emit\_ethernet\$ipv6\_icmp}   & - \\  \hline
\end{tabular}}}
\end{table}

\begin{figure}[htbp]
\centering
  \includegraphics[width=0.9\linewidth]{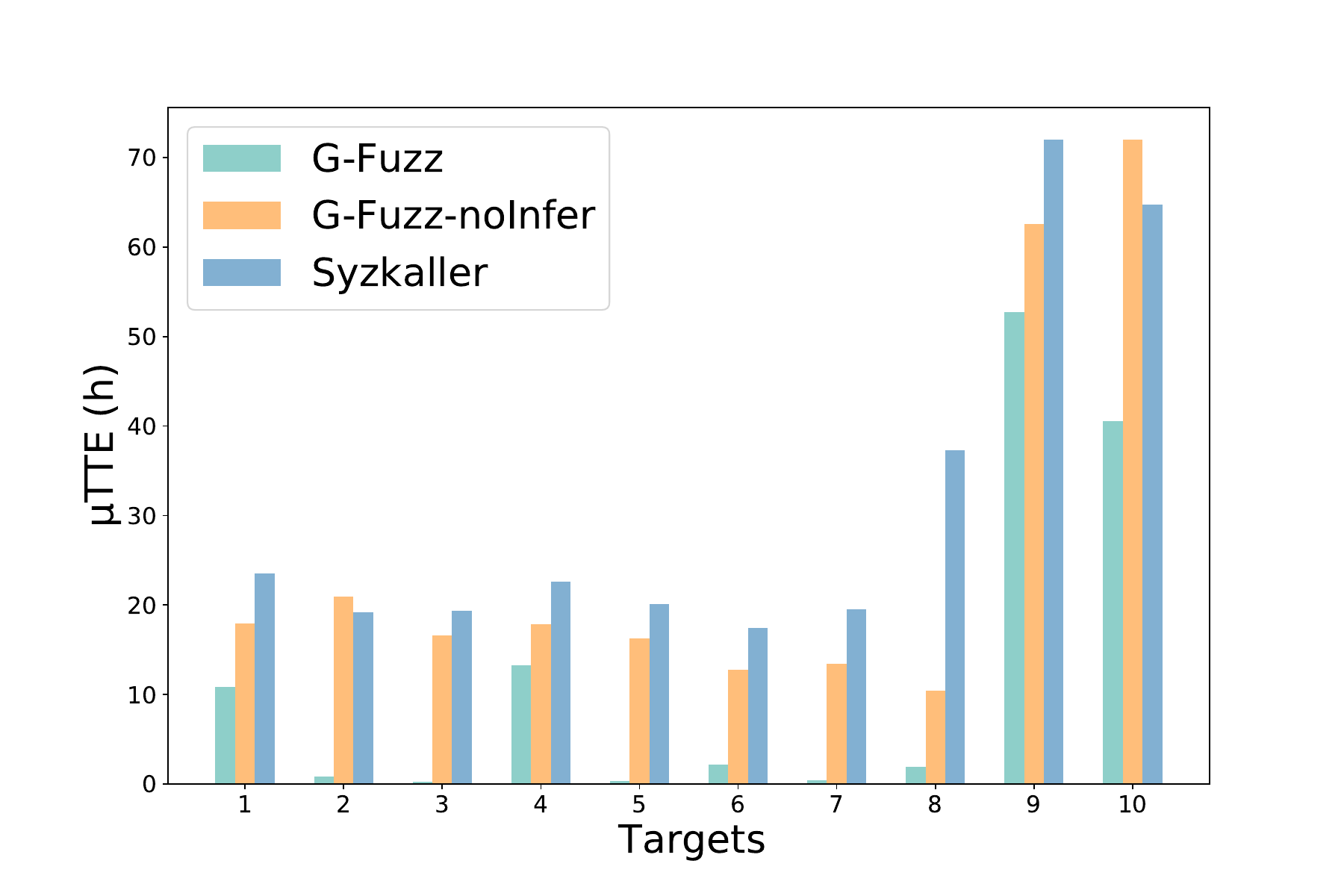}
  \caption{Ablation study 2: target related syscall inference optimization.}
  \label{fig:infer}
   \vspace{-5mm}
\end{figure}

{\color{black}
\textbf{Ablation Study.}
We conduct ablation study on the target related syscall inference method.
we compare G-Fuzz to G-Fuzz-noInfer (G-Fuzz that does not leverage the inferred syscalls) and Syzkaller.
Fig.~\ref{fig:infer} shows the experimental results, we can obverse that G-Fuzz uses significantly less time in triggering the targets than G-Fuzz-noInfer and Syzkaller.
G-Fuzz outperforms G-Fuzz-noInfer and Syzkaller in 10 targets.
G-Fuzz-noInfer outperforms Syzkaller in 8 targets (except for target 2 and 10).
The results can prove the effectiveness of the target inference optimization of G-Fuzz.
}

\subsection{Exploration and Exploitation Dynamic Switch}

\textbf{The Selection of Parameters.}
{\color{black}
To test the impacts of the selection of parameters (i.e., $T_{a}$ and $T_{b}$), we conduct the experiments.
Specifically, we select 10 targets to test the performance of G-Fuzz with different parameter settings.
Each experiment is conducted in five repetitions.
As the value space of the two parameters ($T_{a}, T_{b}$) is very large, for convenience, we set $T_{a}=T_{b}$ in the experiments.
Table~\ref{table:tatb} shows the experimental results.
The second column represents the experimental results of G-Fuzz with the setting in this paper ($T_{a}=5min, T_{b}=10min$).
From the table, we can observe that in most cases, G-Fuzz has the best performance with the paper's setting.
On two targets, G-Fuzz with the setting ($T_{a}=T_{b}=1min$) has the best performance.
If the value of $T_{a}$ and $T_{b}$ is set too large, it may hinder the performance of G-Fuzz.
}

\begin{table*}[!tbp]
\centering
\footnotesize
\caption{The performance of G-Fuzz with different $T_{a}$ and $T_{b}$ settings.}
\label{table:tatb}
\setlength{\tabcolsep}{10pt}{
\scalebox{0.9}{
\begin{tabular}{c|cccccc|c}
\hline
\multirow{2}{*}{Targets} & \multicolumn{6}{c|}{G-Fuzz}       & \multirow{2}{*}{Syzkaller} \\ \cline{2-7}
                         & \multicolumn{1}{c|}{paper-setting}  & \multicolumn{1}{c|}{$T_{a}=T_{b}$=1min} & \multicolumn{1}{c|}{$T_{a}=T_{b}$=30min} & \multicolumn{1}{c|}{$T_{a}=T_{b}$=1h} & \multicolumn{1}{c|}{$T_{a}=T_{b}$=4h} & $T_{a}=T_{b}$=8h &                            \\ \hline
1                        & \multicolumn{1}{c|}{\textbf{10.81}} & \multicolumn{1}{c|}{18.79}               & \multicolumn{1}{c|}{16.86}               & \multicolumn{1}{c|}{11.28}            & \multicolumn{1}{c|}{13.32}            & 15.03            & 23.52                      \\ \hline
2                        & \multicolumn{1}{c|}{\textbf{0.80}}  & \multicolumn{1}{c|}{9.90}                & \multicolumn{1}{c|}{15.85}               & \multicolumn{1}{c|}{7.03}             & \multicolumn{1}{c|}{12.00}            & 8.73             & 19.17                      \\ \hline
3                        & \multicolumn{1}{c|}{0.23}           & \multicolumn{1}{c|}{\textbf{0.14}}       & \multicolumn{1}{c|}{9.70}                & \multicolumn{1}{c|}{0.22}             & \multicolumn{1}{c|}{4.98}             & 0.16             & 19.39                      \\ \hline
4                        & \multicolumn{1}{c|}{13.28}          & \multicolumn{1}{c|}{17.36}               & \multicolumn{1}{c|}{17.49}               & \multicolumn{1}{c|}{20.65}            & \multicolumn{1}{c|}{22.71}            & 18.10            & 22.62                      \\ \hline
5                        & \multicolumn{1}{c|}{\textbf{0.35}}  & \multicolumn{1}{c|}{5.24}                & \multicolumn{1}{c|}{1.59}                & \multicolumn{1}{c|}{5.29}             & \multicolumn{1}{c|}{3.13}             & 12.01            & 20.11                      \\ \hline
6                        & \multicolumn{1}{c|}{2.21}           & \multicolumn{1}{c|}{\textbf{0.09}}       & \multicolumn{1}{c|}{0.15}                & \multicolumn{1}{c|}{1.35}             & \multicolumn{1}{c|}{5.32}             & 5.30             & 17.41                      \\ \hline
7                        & \multicolumn{1}{c|}{\textbf{0.42}}  & \multicolumn{1}{c|}{0.71}                & \multicolumn{1}{c|}{11.14}               & \multicolumn{1}{c|}{0.95}             & \multicolumn{1}{c|}{0.85}             & 5.53             & 19.50                      \\ \hline
8                        & \multicolumn{1}{c|}{\textbf{1.90}}  & \multicolumn{1}{c|}{4.57}                & \multicolumn{1}{c|}{4.91}                & \multicolumn{1}{c|}{5.31}             & \multicolumn{1}{c|}{72.00}            & 72.00            & 37.29                      \\ \hline
9                        & \multicolumn{1}{c|}{\textbf{52.77}} & \multicolumn{1}{c|}{54.37}               & \multicolumn{1}{c|}{57.49}               & \multicolumn{1}{c|}{61.60}            & \multicolumn{1}{c|}{72.00}            & 72.00            & 72.00                      \\ \hline
10                       & \multicolumn{1}{c|}{\textbf{40.57}} & \multicolumn{1}{c|}{23.30}               & \multicolumn{1}{c|}{30.18}               & \multicolumn{1}{c|}{29.27}            & \multicolumn{1}{c|}{72.00}            & 72.00            & 64.74                      \\ \hline
\end{tabular}}}
\end{table*}

\begin{table*}[tbp]
\centering
\footnotesize
\caption{The new vulnerabilities of gVisor-x found by G-Fuzz.}
\label{tab:gfuzz-realworld}
\setlength{\tabcolsep}{10pt}{
\scalebox{0.9}{
\begin{tabular}{l|l}
\hline
 Type             & Description                                                                           \\ \hline
 buffer overflow      & panic: runtime error: slice bounds out of range {{[}}12:LINE{{]}}                 \\ \hline
buffer overflow      & panic: runtime error: index out of range {{[}}260{{]}} with length 260            \\ \hline
stack overflow       & fatal error: heapBitsSetType: called with non-pointer type                            \\ \hline
logic error          & panic: Unknown syscall NUM error: rename across inodes with different implementations \\ \hline
floating point error & panic: runtime error: floating point error                                            \\ \hline
concurrency bug      & fatal error: s.freeindex \textgreater s.nelems                                        \\ \hline
concurrency bug      & panic: gofer.dentry.decRefNoCaching() called without holding a reference              \\ \hline
\end{tabular}}}
\end{table*}

\begin{figure}[htbp]
\centering
  \includegraphics[width=0.9\linewidth]{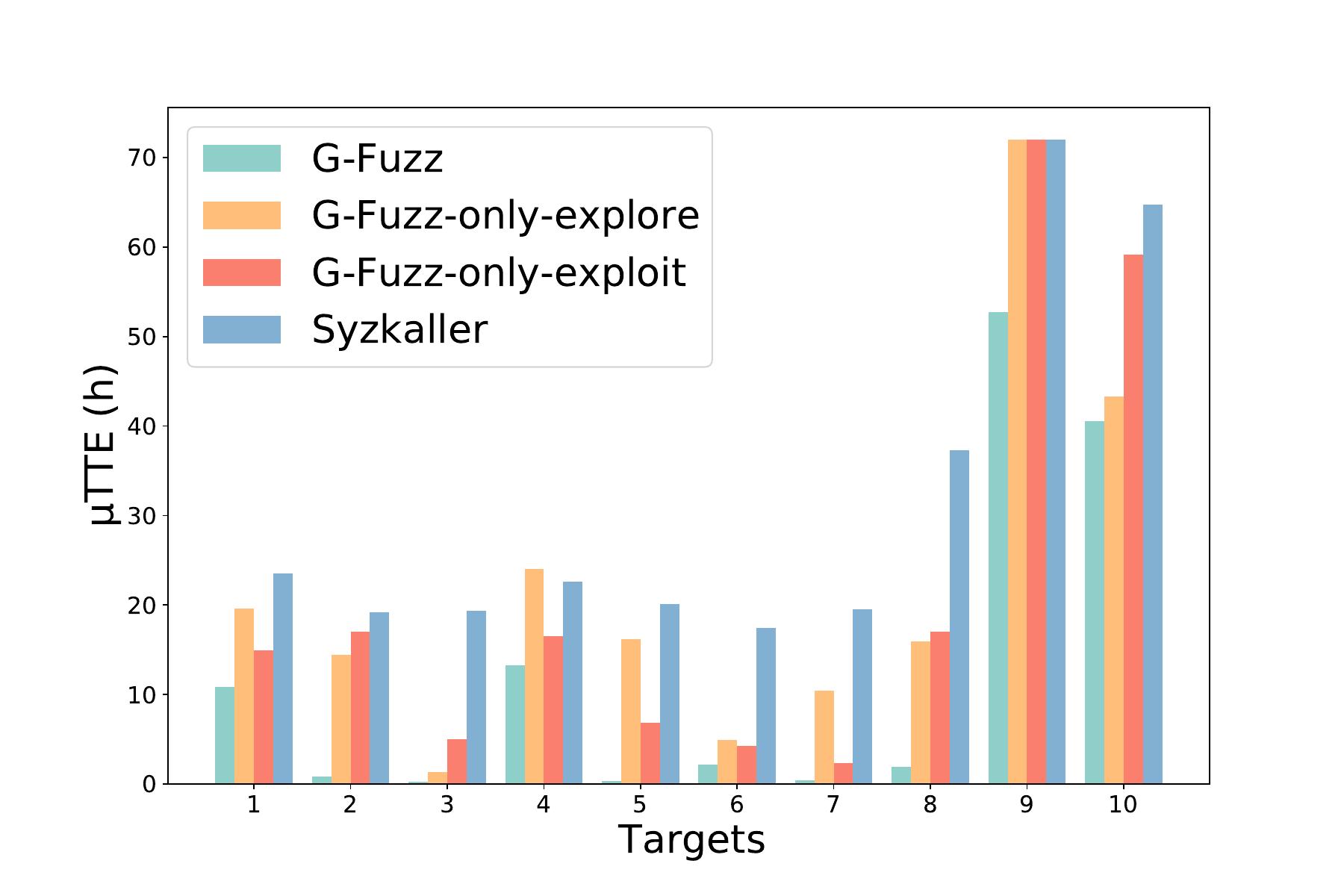}
  \caption{Ablation study 3: exploration and exploitation dynamic switch optimization.}
  \label{fig:switch}
   \vspace{-5mm}
\end{figure}

{\color{black}
\textbf{Ablation Study.}
We conduct ablation study on the exploration and exploitation dynamic method.
we compare G-Fuzz to G-Fuzz-only-explore (G-Fuzz that only uses exploration state), G-Fuzz-only-exploit (G-Fuzz that only uses exploitation state) and Syzkaller.
Fig.~\ref{fig:switch} shows the experimental results.
We can find that G-Fuzz has the best performance, which proves the effectiveness of the exploration and exploitation dynamic switch optimization.
As G-Fuzz-only-explore has better performance than G-Fuzz-only-exploit in 4 out of 10 targets (target 2, 3,  8, 10), it is hard to determine which is better.
In total, G-Fuzz-only-explore and G-Fuzz-only-exploit use less time in triggering targets than Syzkaller, which also demonstrate the powers of the other optimizations of G-Fuzz in directed fuzzing.
}

\subsection{The Deployment and the Application of \sys}

We have applied \sys in directly fuzzing the customized \gv, called gVisor-x, which is widely deployed in Ant Group, a world-leading IT company.
We select several vulnerable code from Syzbot~\cite{syzbot-gVisor}, and if the vulnerable code lies in gVisor-x, it will be considered as the target to conduct directed fuzzing test.
Leveraging \sys, we successfully discovered seven new vulnerabilities in gVisor-x, which the developers of Ant Group have confirmed.
For commercial reasons, we present partial information of the discovered 
new vulnerabilities in Table~\ref{tab:gfuzz-realworld}.
The found vulnerabilities may have severe impacts on the security and reliability of the business service. 
For instance, a malicious process inside the container can make the \gv panic, so the business process will also be killed, causing Deny of Service.
Moreover, the race condition vulnerability may potentially allow the attacker to gain privilege.

\subsection{The Generalisability of \sys}
{\color{black}
Despite we focus on solving the directed fuzzing problem for \gv in this paper. 
The three core methods of \sys are general and scalable, which address the critical challenges for applying directed fuzzing on other OS kernels.
First, the static distance calculation of \sys can be applied to any program that can be compiled into LLVM IR.
Second, most of the inference rules for identifying the target related syscalls are suitable for other OS kernels.
Third, the exploration and exploitable dynamic switch method can be applied to directed fuzzing for any program.
Moreover, to prove the scalability of \sys, we implement a simple directed fuzzing prototype for Linux kernels called G-Fuzz-Linux, which is based on the methods of \sys.  
We conduct experiments to compare the performance of G-Fuzz-Linux with Syzkaller in directed fuzzing Linux kernels.
Here we select ten Linux kernel bugs as the targets and repeat each experiment in five times.
Table~\ref{tab:g-fuzz-linux} presents the experimental results.
We can observe that G-Fuzz-Linux has better performance than Syzkaller in reproducing Linux kernel bugs.
Bug 3, 4, and 10 can only be triggered by G-Fuzz-Linux.
Moreover, we open source the code of G-Fuzz-Linux~\cite{gfuzz-linux} to facilitate the future research.
}

\begin{table}[!tbp]
\centering
\footnotesize
\caption{The performance of G-Fuzz-Linux and Syzkaller in Linux kernel bug reproduction.}
\label{tab:g-fuzz-linux}
\setlength{\tabcolsep}{10pt}{
\scalebox{0.9}{
\begin{tabular}{c|l|c|c|c}
\hline
Bug                 & Fuzzer       & Runs & $\mu$TTE (h) & Version                   \\ \hline
\multirow{2}{*}{1}  & G-Fuzz-Linux & 5    & 4.06         & \multirow{6}{*}{4.19.204} \\ \cline{2-4}
                    & Syzkaller    & 1    & 65.52        &                           \\ \cline{1-4}
\multirow{2}{*}{2}  & G-Fuzz-Linux & 5    & 4.22         &                           \\ \cline{2-4}
                    & Syzkaller    & 1    & 71.17        &                           \\ \cline{1-4}
\multirow{2}{*}{4}  & G-Fuzz-Linux & 3    & 60.79        &                           \\ \cline{2-4}
                    & Syzkaller    & 0    & 72.00        &                           \\ \hline
\multirow{2}{*}{3}  & G-Fuzz-Linux & 1    & 61.40        & \multirow{4}{*}{4.19.199} \\ \cline{2-4}
                    & Syzkaller    & 0    & 72.00        &                           \\ \cline{1-4}
\multirow{2}{*}{6}  & G-Fuzz-Linux & 5    & 18.27        &                           \\ \cline{2-4}
                    & Syzkaller    & 1    & 70.91        &                           \\ \hline
\multirow{2}{*}{7}  & G-Fuzz-Linux & 5    & 36.71        & \multirow{2}{*}{4.19.194} \\ \cline{2-4}
                    & Syzkaller    & 1    & 68.49        &                           \\ \hline
\multirow{2}{*}{5}  & G-Fuzz-Linux & 5    & 38.59        & \multirow{4}{*}{4.19.186} \\ \cline{2-4}
                    & Syzkaller    & 1    & 58.64        &                           \\ \cline{1-4}
\multirow{2}{*}{8}  & G-Fuzz-Linux & 5    & 40.77        &                           \\ \cline{2-4}
                    & Syzkaller    & 1    & 70.89        &                           \\ \hline
\multirow{2}{*}{9}  & G-Fuzz-Linux & 5    & 6.39         & \multirow{4}{*}{4.19.180} \\ \cline{2-4}
                    & Syzkaller    & 1    & 59.16        &                           \\ \cline{1-4}
\multirow{2}{*}{10} & G-Fuzz-Linux & 3    & 37.92        &                           \\ \cline{2-4}
                    & Syzkaller    & 0    & 72.00        &                           \\ \hline
\end{tabular}}}
\end{table}

\section{Discussion: Threats to Validity}

Although \sys has good performance in directed fuzzing for \gv, it can be further improved in the following ways.

\textbf{Optimization of the Inference Methods.}
The proposed target related syscall inference method creatively reduces the huge input space for kernel directed fuzzing.
The experimental results also prove the effectiveness of this method.
The existing inference methods can be improved from the following ways.
First, the existing methods are based on the expert knowledge.
Some inference rules are easily to generate by analyzing the target code and the PoC input, while some rules still need much expert knowledge.
In essence, the inference rules reflect the relations of the properties of the target code and the PoC inputs.
Thus, it is feasible to develop an automatic inference rule generation method from large-scale corpus with machine learning algorithms in the future.
Second, the inference rules are generated by static analysis.
In the future, we will study how to generate the inference rules dynamically during the fuzzing process.

\textbf{Discovery of More Vulnerabilities.}
Due to the limitation of the targets and the fuzzing time, we do not find a large amount of new vulnerabilities in the current paper. 
As G-Fuzz has outstanding performance in directed fuzzing, which is also demonstrated from the existing experimental results, we are confident that G-Fuzz can discover new vulnerabilities when testing more targets.

\section{Related Work}
\textbf{Kernel Fuzzing.}
Fuzzing is a popular research topic in software and system security~\cite{mopt, unifuzz-li, Parallel-zhang, vfuzz-li}.
There are many research works focus on kernel fuzzing~\cite{han2017imf, kAFL, pailoor2018moonshine,songperiscope,corina2017difuze,peng2020usbfuzz,xufuzzing,kim2020finding,jeong2018razzer,xu2020krace}.
IMF~\cite{han2017imf} is a model-based API fuzzer that targets macOS.
Syzkaller~\cite{Syzkaller} is a widely used kernel fuzzer and can be used for multiple OS kernels such as Linux, OpenBSD, Windows, gVisor, etc.
kAFL~\cite{kAFL} proposes a hardware-assisted method to record coverage with low overhead, and can be applied to many OS kernels.
Moonshine~\cite{pailoor2018moonshine} leverages static analysis to extract the dependencies across different syscalls to further distilling the initial seeds for OS kernel fuzzing.
In \sys, we also analyze the dependencies of the syscalls to generate semantically correct seeds.
Some OS kernel fuzzers aim at drivers~\cite{songperiscope,corina2017difuze,peng2020usbfuzz}.
DIFUZE~\cite{corina2017difuze} is an interface aware fuzzer for Android device drivers.
Moreover, some kernel fuzzing works aim at testing the file system of the kernel~\cite{xufuzzing,kim2020finding} or detecting some specific vulnerabilities~\cite{jeong2018razzer,xu2020krace}.

\textbf{Static Analysis for Kernel.}
A plethora of research has adopted static analysis to discover different vulnerabilities of the OS kernels~\cite{gens2018k, xu2018precise, luo2018dftinker, wang2017double, bai2019dcns, wang2018check, lu2019detecting}.
K-Miner~\cite{gens2018k} is a framework for detecting memory corruption related vulnerabilities of the Linux kernel.
Deadline~\cite{xu2018precise}, Dftinker~\cite{luo2018dftinker} and \cite{wang2017double} focus on discovering double-fetch vulnerabilities in the Linux kernel.
DCNS~\cite{bai2019dcns} aims at detecting non-sleep defects in the Linux kernel.
For missing security checks, LRSan~\cite{wang2018check} and Crix~\cite{lu2019detecting} are proposed for detecting such vulnerabilities.  
The kernel code is complex and has many indirect calls, which brings challenges to the static analysis.
Lu et al.~\cite{lu2019does} propose multi-layer type analysis to identify the indirect calls with outstanding performance, demonstrating the effectiveness of type analysis methods.
In this paper, we also leverage type analysis to find the indirect calls of \gv.

\section{Conclusion}
We propose \sys, a directed fuzzing framework  for \gv, which incorporates three core methods including \emph{lightweight and fine-grained distance calculation},
\emph{target related syscalls inference and utilization},
and
\emph{exploration and exploitation dynamic switch}.
The methods of \sys are general and scalable, which can be applied on other OS kernels like Linux kernels.
Compared to the state-of-the-art kernel fuzzers, \sys achieves much more efficient and stable performance.
Moreover, \sys has been deployed in the industry for the customized \gv, and has successfully discovered multiple real-world vulnerabilities.

\section*{Acknowledgments}
We sincerely appreciate the reviewers for their valuable comments to improve our paper.
This work is supported by the ``Pioneer" and ``Leading Goose" R\&D Program of Zhejiang (2022C01243), NSFC under No. 62202484, NSFC under No. U1936215, the State Key Laboratory of Computer Architecture (ICT, CAS) under Grant No. CARCHA202001, and the Fundamental Research Funds for the Central Universities (Zhejiang University NGICS Platform).

\bibliographystyle{IEEEtran}

\bibliography{IEEEabrv,main}




\end{document}